\documentclass[journel]{IEEEtran}
\pdfoutput=1

\usepackage{epsfig}
\usepackage[usenames]{color}
\usepackage{amsmath}
\usepackage{amssymb}
\usepackage{mathrsfs}
\usepackage{amsfonts}
\usepackage{lscape}
\usepackage{graphicx}

\newcommand{\Xcal}{\mathcal{X}}

\newcommand{\X}{X}

\newcommand{\sign}{\hbox{sign}}

\newcommand{{\main}}{\hbox{\textsl{main}}}

\newcommand{\E}{\mathbb{E}}

\newcommand{\PP}{\mathbb{P}}

\newcommand{\intersect}{\cap}
\newcommand{\union}{\cup}

\newcommand{\algg}[1]{\begin{align*} #1 \end{align*}}
\newcommand{\sst}{S^{*}}

\newcommand{\dalp}{\bigtriangleup_{\alpha}}
\newcommand{\dtalp}{\tilde{\bigtriangleup}_{\alpha}}
\newcommand{\tdalp}{\dtalp}

\newcommand{\calph}{C_{\alpha}}

\newcommand{\tts}{\tilde{T}(S)}
\newcommand{\ymxs}{\frac{ |Y - X_S| ^2}{\sigma^2}}
\newcommand{\ymaxst}{\frac{|Y - (1\!-\!\alpha )X_{\sst}|^2}{\sigma^2 + \alpha^2 P}}

\begin{document}

\title{Least Squares Superposition Codes of Moderate Dictionary Size, Reliable 
at Rates up to Capacity}
\author{Andrew~R.~Barron,~\IEEEmembership{Senior~Member,~IEEE,}
        and~Antony~Joseph,~\IEEEmembership{Student~Member,~IEEE}
\thanks{Andrew R. Barron and Antony Joseph
are with the Department of Statistics, Yale University, New Haven, CT 06520 USA
e-mail:  \{andrew.barron,~antony.joseph\}@yale.edu.}
\thanks{Summary \cite{BarronLeast} of this paper was presented at the \textit{IEEE International Symposium on Information Theory}, Austin, Texas, June 13-18, 2010.}
\\Submitted to \textit{IEEE Transactions on Information Theory},~June 4, 2010.}





\maketitle

\begin{abstract}
For the additive white Gaussian noise channel with average codeword power constraint, new coding methods are devised in which the codewords are sparse
superpositions, that is, linear combinations of subsets of vectors from a given design, with the possible messages indexed by the choice of subset.
Decoding is by least squares, tailored to the assumed form of linear combination.  Communication is shown to be reliable with error probability exponentially small for all rates up to the Shannon capacity.
\end{abstract}

\section{Introduction}
\label{sec:intro}
The additive white Gaussian noise channel is basic to Shannon theory and underlies practical communication models.  We introduce classes of superposition codes for this channel and analyze their properties.  We link theory and practice by showing superposition codes from polynomial size dictionaries with least squares decoding achieve exponentially small error probability for any communication rate 
less than the Shannon capacity.  A companion paper \cite{BarronJosephFast},\cite{BarronJosephFastC} provides a fast decoding method and its analysis. The developments involve a merging of modern perspectives on statistical linear model selection and information theory.

The familiar 
communication problem is as follows.  An encoder is required to map input bit strings $u=(u_1,u_2,\ldots,u_K)$ of length $K$ into codewords which are length $n$ strings of real numbers $c_1,c_2,\ldots,c_n$, with 
norm expressed 
via the power $(1/n)\sum_{i=1}^n c_i^2$.
We constrain the average of the power across the $2^K$ codewords to be not more than $P$.
The channel adds independent $N(0,\sigma^2)$ noise to the selected codeword yielding a received length $n$ string 
$Y$. A decoder is required to map 
it into an estimate $\hat u$ which we want
to be a correct decoding of 
$u$.  Block error is the event $\hat u \neq u$,
bit error at position $i$ is the event $\hat u_i \neq u_i$, and 
the bit error rate is
$(1/K)\sum_{i=1}^K 1_{\{\hat u_i \neq u_i\}}$. An analogous section error rate for our code is defined below.
The reliability requirement is that, with sufficiently large $n$, the bit error rate or section error rate is small with high probability or, more stringently, 
the block error probability is small, 
averaged over 
input strings $u$ as well as the distribution of $Y$. The communication rate $R = K/n$ is the ratio of 
the input length to the codelength for communication across the channel. 

The supremum 
of reliable rates 
 is the channel capacity $C \!=\! (1/2) \log_2 (1 \!+\! P/\sigma^2)$, 
by traditional information theory 
as
in  \cite{Shannon1948}, \cite{Gallager1968}, \cite{CoverThomas2006}.
Standard communication models, even in continuous-time, have been reduced to the above discrete-time
white Gaussian noise setting, as in \cite{Gallager1968},\cite{ForneyAWGN}.
This problem is also of interest in mathematics because of 
relationship to versions of the sphere packing problem as described in Conway and Sloane \cite{ConwaySloane1988}. For practical coding the challenge is to achieve 
rates arbitrarily close to capacity with a codebook of moderate size,
while guaranteeing 
reliable decoding in manageable computation time.

We introduce a new coding scheme based on sparse superpositions with a moderate size dictionary and analyze its performance. Least squares is the optimal decoder. Accordingly, we analyze the reliability of least squares and approximate least squares decoders. The analysis here is without concern for computational feasibility. In similar settings computational feasibility is addressed in the companion paper \cite{BarronJosephFast},\cite{BarronJosephFastC}, though the closeness to capacity at given reliability levels is not as good as developed here.


We introduce sparse superposition codes and discuss the reliability of least squares in Subsection \ref{sub:spar} of this Introduction. Subsection \ref{sub:decod} contrasts the performance of least squares with what is achieved by other methods of decoding. In Subsection \ref{sub:pracd}, 
we mention relations with work on sparse signal recovery in the high dimensional regression setting.
Subsection \ref{sub:AWGNcode} discusses other codes and Subsection \ref{sub:forneycover} discusses some important forerunners to our developments here. Our reliability bounds are
developed in subsequent sections.

\subsection{Sparse Superposition Codes}
\label{sub:spar}
We develop the framework for code construction by linear combinations.  The story begins with a list (or book) $\X_1,\X_2,\ldots,\X_N$ of vectors, each with $n$ coordinates, for which the codeword vectors take the form of superpositions
$\beta_1 \X_1 + \beta_2 \X_2 + \ldots + \beta_N \X_N$.  The vectors $\X_j$ which are linearly combined provide the terms or components of the codewords and the $\beta_j$ are the coefficients. The received vector is in accordance with the statistical linear model $$Y=X\beta+ \varepsilon$$ where $X$ is the matrix whose columns are the vectors $\X_1,\X_2,\ldots,\X_N$ and $\varepsilon$ is the noise vector distributed Normal($0,\sigma^2 I$). In keeping with the terminology of that statistical setting, the book $X$ may be called the design matrix consisting of $p=N$ variables, each with $n$ observations, and this list of variables is also called the dictionary of candidate terms.

The coefficient vectors $\beta$ are arranged to be of a specified form.
For \textit{subset superposition coding} we arrange for a number $L$ of the coordinates to be non-zero, with a specified positive value, and the message is conveyed by the choice of subset.  Denote $B=N/L$.  
If $B$ is large, it is a \textit{sparse superposition code}. In this case, the number of terms sent is a small fraction
of dictionary size.
 With somewhat greater freedom, 
one may arrange the non-zero coefficients to be $+1$ or $-1$ times a specified value, in which case the superposition code is said to be \textit{signed}. Then the message is conveyed by the sequence of signs as well as the choice of subset.

To allow such forms of $\beta$, we do not in general take the set of permitted coefficient vectors to be closed under a field of linear operations, and hence our linear statistical model does not correspond to a linear code in the sense of traditional algebraic coding theory.  

In a specialization we call a \textit{partitioned superposition code}, the book $X$ is split into $L$ sections of 
size $B$, with one term 
selected from each, yielding $L$ terms in each codeword out of a dictionary of size  $N\!=\!LB$.  
Likewise, the coefficient vector $\beta$ is split into sections, 
with one coordinate 
non-zero in each section to indicate the selected term. Optionally, we have the additional freedom of choice of sign of this coefficient, for 
a signed partitioned code.
It is desirable that the section sizes be 
not larger than a moderate order polynomial in $L$ or $n$, for then the dictionary is arranged to be of manageable size.

Most convenient is the case that the sizes of these sections are powers of two. Then 
an input bit string of length $K\!=\! L \log_2 B$ splits into $L$ substrings of size $\log_2 B$. The encoder mapping from $u$ to $\beta$ is then obtained by interpreting each substring of $u$ as simply giving the index of which coordinate of $\beta$ is non-zero in the corresponding section. That is, each substring is the binary representation of the corresponding index.

As we have said, the rate of the code is $R=K/n$ input bits per channel uses and we arrange for $R$ arbitrarily close to $C$. For the partitioned superposition code, this rate is $R = (L\log B)/n$. For specified rate $R$, the codelength
$n= (L/R) \log B$. Thus, the length $n$ and the number of terms $L$ agree to within a log factor.

With one term from each section, the number of possible codewords $2^K$ is equal to $B^L = (N/L)^L$. 
Alternatively, if we allow for all subsets of size $L$, the number of possible codewords would be $ N \choose L $, which is of order $(Ne/L)^L = (Be)^L$, for $L$ small compared to $N$. To match the number of codewords, it would correspond to
reducing $N$ by a factor of $1/e$.
Though there would be the factor $1/e$ savings in dictionary size from allowing all subsets of the specified size, the additional simplicity of implementation and simplicity of analysis with partitioned coding is such that 
we take advantage of it wherever appropriate.

With signed partitioned coding the story is similar, now with $(2B)^L= (2N/L)^L$ possible codewords using the dictionary of size $N=LB$.  The input string of length $K= L \log_2 (2B)=L(1+\log_2 B)$, splits into $L$ sections 
with $\log_2 B$ bits to specify the non-zero term and $1$ bit to specify its sign.  For a rate $R$ code this entails a codelength of $n=(L/R) \log (2B)$.  

Control of the 
dictionary size is critical to computationally advantageous coding and decoding. Possible dictionary sizes are between the extremes $K$ and $2^K$ dictated by the number and size of the sections, where $K$ is the number of input bits. At one extreme, with $1$ section of size $B\!=\!2^K$, one has $X$ as the whole codebook with its columns as the codewords,
but the exponential size 
makes its direct use impractical.  At the other extreme we have $L\!=\!K$ sections, each with 
two candidate terms in subset coding 
or two signs of a single term in sign coding with $B\!=\!1$;
in which case $X$ is the generator matrix of a linear code.

Between these extremes, we construct 
reliable, high-rate codes with
codewords corresponding to linear combinations of subsets of terms in moderate size dictionaries.

Design of the dictionary is guided by what is known from information theory concerning the distribution of symbols in the codewords.
By analysis of the converse to the channel coding theorem (as in \cite{CoverThomas2006}), for a reliable code at rate near capacity, with a uniform distribution on the sequence of input bits, the induced empirical distribution on coordinates of the codeword must be close to independent Gaussian, in the sense that the resulting mutual information must be close to its maximum subject to the power constraint.

We draw entries of $X$ independently from a normal distribution with mean zero and a variance we specify,
yielding the properties we want with high probability.
Other distributions, such as independent equiprobable $\pm 1$, might also suffice, with a near Gaussian shape for the codeword distribution obtained by the convolutions associated with sums of terms in subsets of size $L$.


For the vectors $\beta$, the 
non-zero coefficients may be assigned to have magnitude $\sqrt {P/L}$, which with $X$ having independent entries of variance $1$, yields codewords $X\beta$ of average power near $P$.
There is a freedom of scale that allows us to simplify the coefficient representation.  Henceforth, we arrange the 
coordinates of $\X_j$ to have variance $P/L$ and set the non-zero coefficients to have magnitude $1$.  

Optimal decoding for minimal average probability of error consists of finding the codeword $X\beta$ with coefficient vector $\beta$ of the assumed form that maximizes the posterior probability, conditioning on $X$ and $Y$.  This coincides, in the case of equal prior probabilities, with the maximum likelihood rule of seeking such a codeword to minimize the sum of squared errors in fit to $Y$.  This is a least squares regression problem $\min_\beta \|Y \! - \! X\beta\|^2$, with constraints on the coefficient vector.

\begin{figure*}
\centerline{
\mbox{
\includegraphics[width=3.00in]{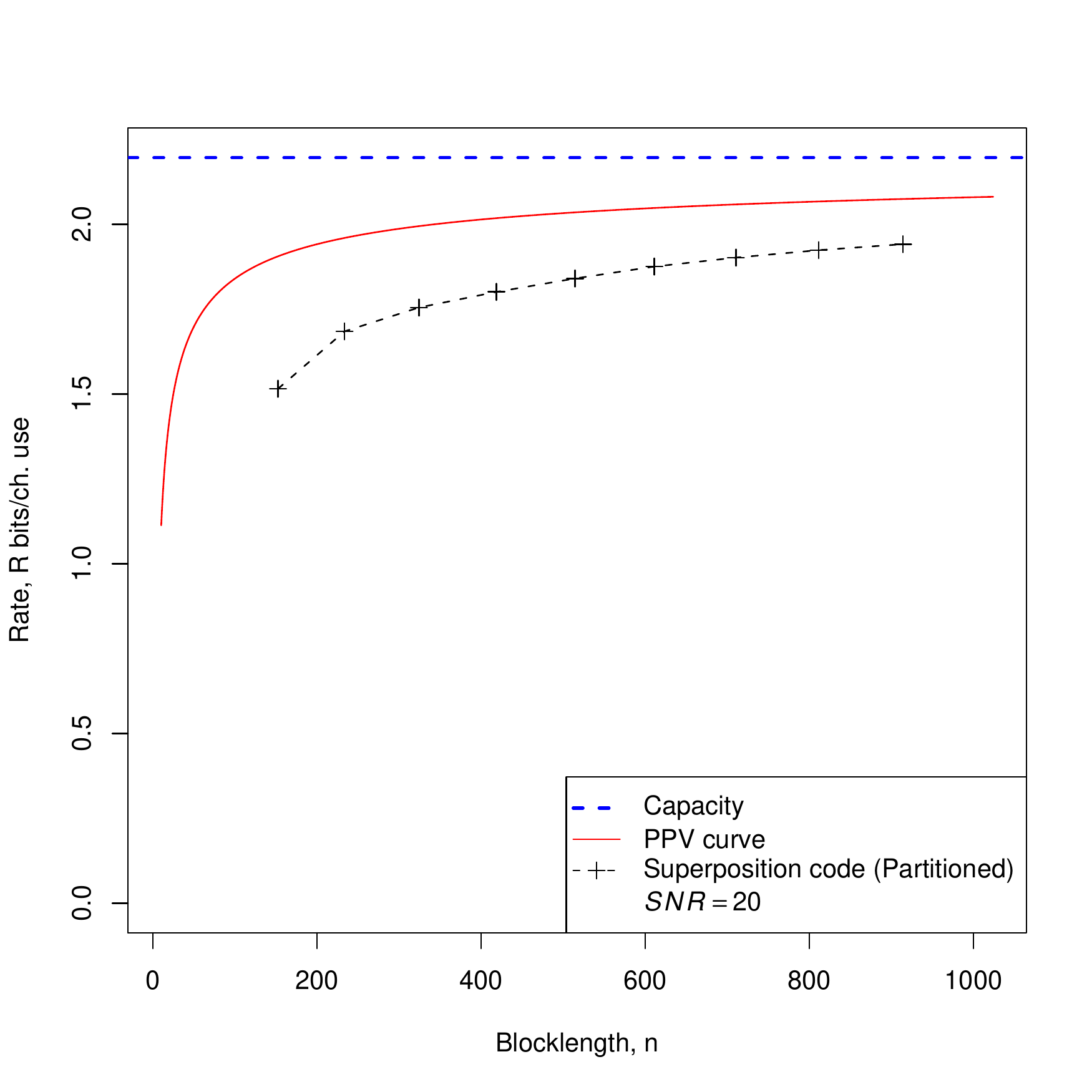}
}
\mbox{
\includegraphics[width=3.00in]{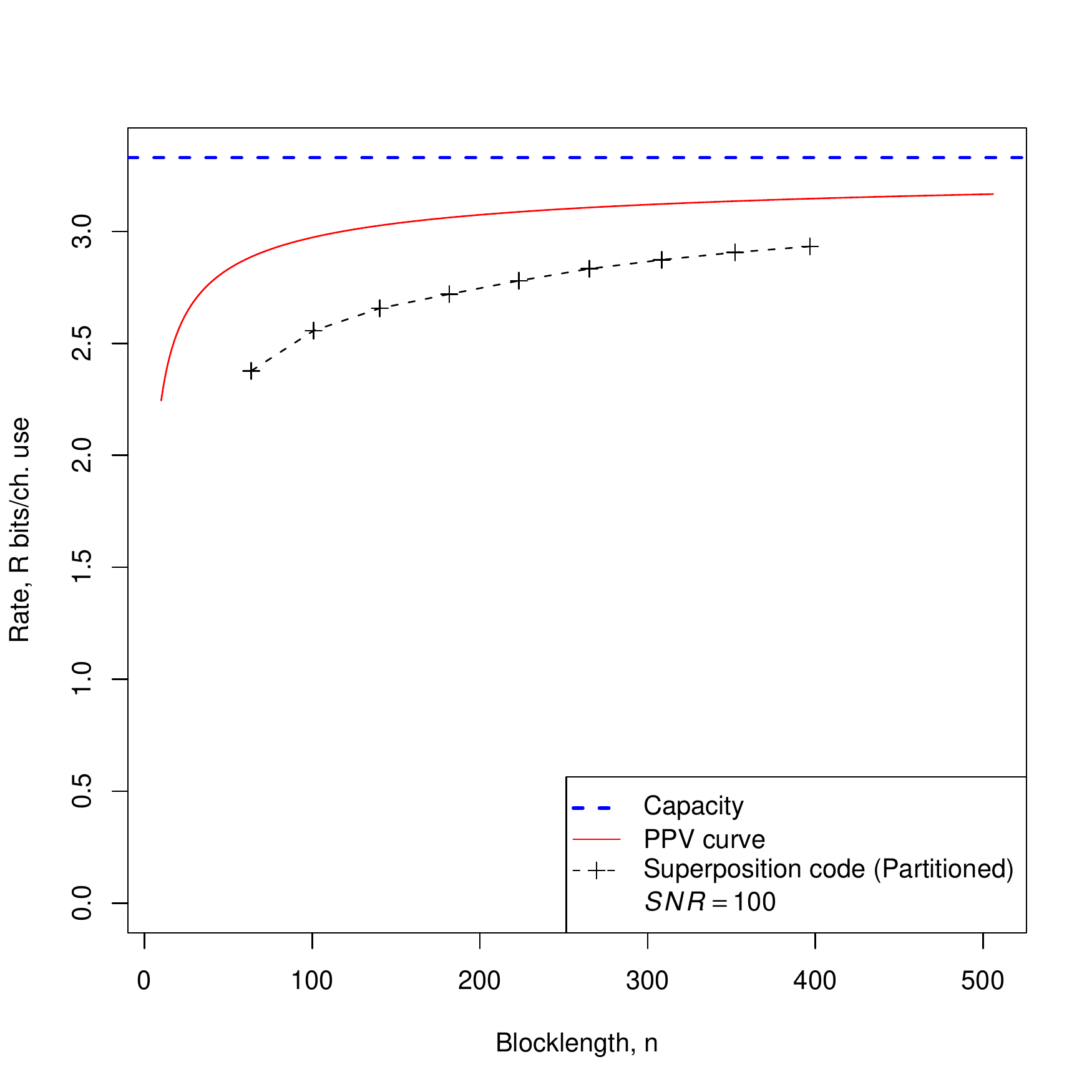}
}
}
\caption{Plot of comparison between achievable rates using our scheme and the theoretical best possible rates for block error probability of $10^{-4}$ and signal-to-noise ratio ($v$) values of 20 and 100. The curves for our partitioned
superposition code were evaluated at points with number of sections $L$  ranging from 20 to 100 in steps of 10, with corresponding $B$ values taken to be $L^{a_v}$, where $a_v$ is as given in Lemma 4 later on. For the $v$ values of $20$ and $100$ shown above, $a_v$ is around $2.6$ and $1.6$, respectively.
}
\label{fig:achrate}
\end{figure*}


We show for all $R < C$,
that the least squares solution, as well as approximate least squares solutions such as may arise computationally, will have, with high probability, at most a negligible fraction of terms that are not correctly identified, producing a low bit error rate.  The heart of the analysis shows that competing codewords that differ in a fraction of at least $\alpha_0$ terms are exponentially unlikely to have smaller distance from $Y$ than the true codeword, provided that the section size $B=L^a$ is polynomially large in the number of sections $L$, where a sufficient value of $a$ is determined.  For the partitioned superposition code there is a positive constant $c$ such that for rates $R$ less than the capacity $C$, with a positive gap $\Delta = C\!-\!R$ not too large, the probability of a fraction of mistakes at least $\alpha_0$ is not more than
$$\exp\{ -nc\min\{\Delta^2,\alpha_0\}\}.$$ 
Consequently, for a target fraction of mistakes $\alpha_0$ and target probability $\epsilon$, the required number of sections $L$ or equivalently the codelength $n=(a L \log L)/R$ depends only polynomially on the reciprocal of the gap $\Delta$ 
and on the reciprocal of $\alpha_0$. Indeed $n$ of order $[(1/\alpha_0)+(1/\Delta)^2 ]\log (1/\epsilon)$ suffices for the probability of the undesirable event to be less than $\epsilon$.

Moreover, an approach is discussed which completes the task of identifying the terms by arranging sufficient distance between the subsets, using composition with an outer Reed-Solomon (RS) code of rate near one.  The Reed-Solomon code is arranged to have an alphabet of size $B$ equal to a power of $2$. It is tailored to the partitioned code by having the RS code symbols specify the terms selected from the sections.  The outer RS code corrects the small fraction of remaining mistakes so that we end up not only with small section error rate but also with small block error probability. If $R_{outer} = 1 - \delta$ is the rate of an RS code, with  $0 < \delta < 1 $, then section error rate less than $\alpha_0$ can be corrected, provided $2\alpha_0 < \delta$. Further, if $R_{inner}$ (or simply $R$) is the
rate associated with our inner (superposition) code, then the total rate after correcting for the remaining mistakes is given by $R_{total} = R_{inner}R_{outer}$. The end result, using our theory for the distribution of the fraction of mistakes of the superposition code, is that the block error probability is 
exponentially small. 
One may regard the composite code as a superposition code in which the subsets are forced to maintain at least a certain minimal separation, so that decoding to within a certain distance from the true subset implies exact decoding.

Particular interest is given to the case that the rate $R$ is made to approach the capacity $C$. Arrange $R = C - \Delta_n$ and $\alpha_0 =  \Delta_n^2$. One may let the rate gap $\Delta_n$ tend to zero (e.g. at a $1/\log n$ rate or any polynomial rate not faster than $1/\sqrt{n}$), then the overall rate $R_{tot} = (1 - 2\alpha_0)(C - \Delta_n)$
continues to have drop from capacity of order $\Delta_n$, with the composite code having block error probability of order
$$\exp\{-nc\,\Delta_n^2\}.$$ The exponent above, of order $(C\!-\! R)^2$ for $R$ near $C$, is in agreement with the form of the optimal reliability bounds as in \cite{Gallager1963}, \cite{poorverdu}, though here our constant $c$ is not demonstrated to be optimal.

In Figure \ref{fig:achrate} we plot curves of achievable rates using our scheme for block error probability
fixed at $10^{-4}$ and signal to noise ratios of $20$ and $100$.
We also compare this to a rate curve  given in Polyanskiy, Poor and Verdu \cite{poorverdu} (the PPV curve), where it is  demonstrated that for a Gaussian channel with signal to noise ratio $v$, the block error probability $\epsilon$, codelength $n$ and rate $R$ with an optimal code can be well approximated by the
following relation,
$$R \approx C - \sqrt{\frac{V}{n}}Q^{-1}(\epsilon) + \frac{1}{2}\frac{\log n}{n} $$
where  $V = (v/2)(v+2)\log^2e/(v+1)^2$
is the channel dispersion and $Q$ is the complementary Gaussian cumulative distribution function.

For the superposition code curve, the y-axis gives the highest $R_{comp}$ 
for which the error probability stays below  $10^{-4}$. These curves are based on the minimum of the bounds obtained by our lemma in Section \ref{sec:rells}. We see for the given $v$ and block error probability values, the achievable rates using our scheme are reasonably close to the theoretically best scheme. Note that the PPV curve was computed with an approach that uses a codebook of size that is exponential in blocklength, whereas our dictionary, of size $LB$, is of considerably smaller size.

\subsection{Contrasting Methods of Decoding}
\label{sub:decod}
As we have said the least squares decoder minimizes $||Y - X\beta||^2$ with constraint on the form of coefficient vector $\beta$. It is unknown whether approximate least squares decoding with  rate $R$ near the capacity $C$ is practical in the equal power case studied here.  Alternative methods include an iterative decoder that we discuss briefly here and convex optimization methods discussed here and in subsection \ref{sub:pracd}.

The practical iterative decoder, for the partitioned superposition code, proposed and analyzed in \cite{BarronJosephFast},\cite{BarronJosephFastC}
is called an \textit{adaptive successive decoder}.
Decoding is broken into multiple steps, with the identification of terms
in a step achieved when the magnitude of the inner product between the corresponding $X_j$'s and a computed residual vector is above a specified threshold. The residual vector for each step being obtained as the difference of $Y$ and the contribution from columns decoded in previous steps.

With a rate that is of order $1/\log B$ below capacity, the error probability attained there is exponentially small
in $L/(\log B)^2$, to within a $\log\log B$ factor. This error exponent is slightly smaller than the optimal
$n/(\log B)^2$, obtained here by the least squares scheme. Moreover, as we saw above, the least squares decoder
achieves the optimal exponent for other orders $\Delta_n$ of drop from capacity.


The sparse superposition codes achieving these performance levels at rates near capacity, by least squares and by adaptive successive decoding are different in an important aspect.  For the present paper, we use a constant power
allocation, with the same power $P/L$ for each term. However in \cite{BarronJosephFast}, to yield rates near capacity
we needed a variable power allocation, achieved by a specific schedule of the non-zero $\beta_j$'s. In contrast, if one
were to use equal power allocation for the decoding scheme in \cite{BarronJosephFast}, then reliable decoding holds only up to a threshold rate $R_{0} =  (1/2)P/(P+\sigma^2)$, which is less than the capacity $C$, with the rate and capacity expressed in nats.

The least squares optimization $\min || Y - X\beta||^2$ is made challenging  by the non-convex constraint that there be a specified number of non-zero coefficients, one in each section. Nevertheless, one can consider decoders
based on projection to the convex hull.
This convex hull consists of the $\beta$ vectors which have sum in each section equal to 1. (With signed coding it becomes the constraint that the $l_1$ norm in each section is bounded by 1.) Geometrically, it provides a convex set of linear combinations in which the codewords are the vertices. Decoding is completed with convex projection by moving to a vertex, e.g. with the largest coefficient value in each section. This is a setting in which we initiated investigations, however, in that preliminary analysis, we found that such $l_1$ constrained quadratic optimization allows for successful decoding only for rates up to $R_{thres}$ for the equal power case.  It is as yet unclear what its reliability properties
would be at rates up to capacity $C$ with variable power.

%
\subsection{Related Work on Sparse Signal Recovery}
\label{sub:pracd}
The conclusions regarding communication rate may be also expressed in the language of \textit{sparse signal recovery}
and \textit{compressed sensing}. A number of terms selected from a dictionary is linearly  combined and subject to noise
in accordance with the linear model framework $Y = X\beta + \epsilon$. Let $N$ be the number of variables and $L$
the number of non-zero terms.
 An issue dealt with by these fields, is the minimal number of observations $n$ sufficient to reliably recover the terms. In our setting, the non-zero values of the coefficients are known and $n$ satisfies the relationship $n = (1/R)\log {N \choose L}$ for general subsets and $n = (1/R) L \log (N/L)$ for the partitioned case. We show that reliable recovery is possible provided $R < C$.

The conclusions here complement recent work on sparse signal recovery \cite{CandesTao2005},\cite{Donoho2006},
\cite{DonohoTanner} in the sparse noise case and \cite{Wainwright2009},\cite{Wainwright2007},\cite{WWR2008},\cite{FRG20061},\cite{Tropp2006},\cite{CandesPalm2009} in the Gaussian noise case. Connections between signal recovery and channel coding are also highlighted in \cite{Tropp2006}.
A hallmark of work in signal recovery is allowance for greater generality of signal coefficient values. In the regime as treated here, where $N \gg  L$ and where there is a control on the sum of squares of the coefficients as well as a control on the minimum coefficient value, conclusions from this literature take the form that the best $n$ is of the order $L\log(N/L)$, with upper and lower bounds on the constants derived. It is natural to call (the reciprocal of) the best constant, for a given set of allowed signals and given noise distribution, the {\em compressed sensing capacity} or {\em signal recovery capacity}. 

For the converse results in \cite{Wainwright2007},\cite{WWR2008}, Fano's inequality is used to establish constants related to the channel capacity. Refinements of this work can be found in \cite{FRG20061}. Convex projection methods with $l_1$ constraints as in \cite{Wainwright2009},\cite{Tropp2006},\cite{CandesPalm2009}, have been used for achievability results. The same order of performance is achieved by a maximum correlation estimator \cite{FRG20061}. Analysis of constants achieved by least squares is in \cite{Wainwright2007}, \cite{FRG2006}.  The above analysis, when interpreted in our setting, correspond to saying that these schemes have communication rate that is positive, though at least a fixed amount below the channel capacity.  For our setting, a consequence of the result here is that the signal recovery capacity is equal to the {\em channel capacity}.

\subsection{Related Communication Issues and Schemes}
\label{sub:AWGNcode}
The development here is specific to the discrete-time channel for which $Y_i=c_i+\varepsilon_i$ for $i=1,2,\ldots,n$ with real-valued inputs and outputs and with independent Gaussian noise.  
Standard communication models, even in continuous-time, have been reduced to this discrete-time white Gaussian noise setting, or to parallel uses of such, when there is a frequency band constraint for signal modulation and when there is a specified spectrum of noise over that frequency band, as in \cite{Gallager1968}, \cite{ForneyAWGN}.

Standard approaches, as discussed in \cite{ForneyAWGN}, entail a decomposition of the problem
into separate problems of coding and of shaping of a multivariate signal constellation. For the low
signal-to-noise regime, binary codes suffice for communication near capacity and there is no need for shaping. 
There is prior work concerning reliable communications near capacity for certain discrete input channels.  Iterative decoding algorithms based on statistical belief propagation in loopy networks have been {\it empirically} shown in various works to provide reliable and moderately fast decoding at rates near the capacity for such channels, and mathematically proven to provide such properties in certain special cases, such as the binary erasure channel in \cite{LMSS2001a,LMSS2001b}.  These include codes based on low density parity check codes \cite{Gallager1963} and turbo codes \cite{Berrou,McEliece}. See \cite{Rich},\cite{RichU} for some aspects of the state of the art with such techniques.

A different approach to reliable and computationally feasible decoding to achieve the rates possible with restriction to discrete alphabet signaling, is in the work on {\em channel polarization} of Arikan and Telatar \cite{Arikan2009,ArikanTelatar2008}.  They achieve rates up to the mutual information $I$ between a uniform input distribution
and the output of the channel. Error probability is demonstrated there at a level exponentially small in $n^{1/2}$ for any fixed $R \!<\! I$.  In contrast for our codes the error probability is exponentially small in $n(C\!-\!R)^2$ for the least squared decoder and within a log factor of being exponentially small in $n$ for the practical decoder in \cite{BarronJosephFast,BarronJosephFastC}. Moreover, communication is permitted at higher rates beyond that associated with a uniform input distribution. We are aware from personal conversation with Imre Telatar and Emanuel Abbe that they are investigating the extent to which channel polarization can be adapted to Gaussian signaling.

In the high signal-to-noise regime, 
one needs a greater signal alphabet size. As explained in \cite{ForneyAWGN}, along with coding schemes on such alphabets, additional shaping is required in order to be able to achieve rates up to capacity. Here shaping refers to making the codewords vectors approximate a good packing of points on the $n$ dimensional sphere of square radius dictated by the power. An implication is that, marginally and jointly for any subset of codeword coordinates, the set of codewords should have empirical distribution not far from Gaussian. Notice that we build shaping directly into the coding scheme by the superposition strategy yielding codewords following a Gaussian distribution.

Our ideas of sparse superposition coding are adapted to Gaussian vector quantization in, Kontoyiannis, Gitzenis and Rad \cite{KGR}.  Applicability to vector quantization is natural because of the above-mentioned connection between packing and coding.

\subsection{Precursors}
\label{sub:forneycover}
The analysis of concatenated codes in Forney \cite{Forney1966} is an important forerunner to the development we give here.  He identified benefits of an outer Reed-Solomon code paired in theory with an optimal inner code of Shannon-Gallager type and in practice with binary inner codes based on linear combinations of orthogonal terms (for target rates $K/n$ less than $1$ such a basis is available).
The challenge concerning theoretically good inner codes is that the 
number of messages searched is exponentially large in the inner codelength.
Forney made the inner codelength of logarithmic size compared to the outer codelength as a step toward practical solution.  However, caution is required with such a strategy. Suppose the rate of the inner code has only a small drop from capacity, $\Delta\!=\!C\!-\!R$.  For small inner code error probability, the inner codelength must be of order at least $1/\Delta^2$. So with that scheme one has the undesirable consequence that the required outer codelength becomes exponential in $1/\Delta^2$.

For the Gaussian noise channel, our tactic to overcome that difficulty 
uses a  superposition inner code with a polynomial size dictionary. We use inner and outer codelengths that are comparable, with the outer code used to correct errors in a small fraction of the sections of the inner code.
The overall codelength to achieve error probability $\epsilon$ remains of the order $(1/\Delta^2)\log(1/\epsilon)$.




Another point of relationship of this work with other ideas is the problem of multiple comparisons in hypothesis tests. False discovery rate \cite{Benjamini} for a given significance level, rather than exclusively overall error probability is a recent focus in statistical development, appropriate when considering very large numbers of hypotheses as arise with many variables in regression.  Our theory for the distribution of the fraction of incorrectly determined terms (associated with bit error rate rather than block error rate) provides an additional glimpse of what is possible in a regression setting with a large number of subset hypotheses. The work of \cite{James} is a recent example where subset selection within groups (sections) of variables is addressed by extension of false discovery methods.


The idea of superposition coding for Gaussian noise channels began with Cover \cite{Cover1972} in the context of multiple-user channels. In that setting what is sent is a sum of codewords, one for each message.  Here we are putting that idea to use for the original Shannon single-user problem.  
The purpose here of computational feasibility is different from the original multi-user purpose which was identification of the set of achievable rates.  Another connection with that broadcast channel work by Cover is that for such Gaussian channels, the power allocation can be arranged such that messages can be peeled off one at a time by successive decoding.  Related rate splitting and successive decoding for superposition codes are developed for Gaussian multiple-access problems in 
 \cite{CaoYeh2007} and 
 \cite{RimoldiUrbanke2001}, where in some cases to establish such reductions, rate splitting is applied to individual users.  
However, feasibility has been lacking in part due to the absence of demonstration of reliability at high rate with superpositions from polynomial size code designs.



It is an attractive feature of our solution for the single-user channel that it should be amenable to extension to practical solution of the corresponding multi-user channels, namely, the Gaussian multiple access and Gaussian broadcast channel.

Section \ref{sec:prelim} contains brief preliminaries.  Section \ref{sec:rells} provides core lemmas on the reliability of least squares for our superposition codes. Section \ref{sec:sufsec} analyzes the matter of section size sufficient for reliability. Section \ref{sec:conprob} confirms that the probability of more than a small fraction of mistakes is exponentially small. Section \ref{sec:reed} 
discusses properties of the composition of our code with a binary outer code for correction of any remaining small fraction of mistakes. 
The appendix collects some auxiliary matters.

\section{Preliminaries}
\label{sec:prelim}

For vectors $a,b$ of length $n$, let $\|a\|^2$ be the sum of squares of coordinates, let $|a|^2 = (1/n)\sum_{i=1}^n a_i^2$ be the average square and let $a \cdot b = (1/n)\sum_{i=1}^n a_i b_i$ be the associated inner product.  It is a matter of taste, but we find it slightly more convenient to work henceforth with the norm $|a|$ rather than $\|a\|$.

Concerning the base of the logarithm ($\log$) and associated exponential ($\exp$), base $2$ is most suitable for interpretation and base $e$ most suitable for the calculus.  For instance, 
the rate 
$R=(L \log B)/n$ is measured in bits if the log is base $2$ and nats if the log is base $e$.
Typically, conclusions are stated in a manner that can be interpreted to be invariant to the choice of base, and base $e$ is used for convenience in the derivations.

We make repeated use of the following moment generating function and its associated large deviation exponent in constructing bounds on error probabilities. 
If $Z$ and $\tilde Z$ are normal with means equal to $0$, variances equal to $1$, and correlation coefficient $\rho$ then $\E (e^{(\lambda/2)(Z^2-\tilde Z^2)})$ takes the value  $$1/[1-\lambda^2(1 \!-\! \rho^2)]^{1/2}$$ when $\lambda^2 < 1/(1\!-\!\rho^2)$ and infinity otherwise. So the associated cumulant generating function of $(1/2)(Z^2 - \tilde Z^2)$ is $-(1/2)\log(1-\lambda^2(1 \!-\! \rho^2))$, with the understanding that the minus log is replaced by infinity when $\lambda^2$ is at least $1/(1\!-\rho^2)$. 
For positive $\Delta$ we define the quantity $D=D(\Delta,1\!-\!\rho^2)$ given by
$$D= \max_{\lambda \ge 0} \big\{ \lambda \Delta + (1/2) \log (1- \lambda^2(1 \!-\! \rho^2))\big\}.$$

This $D$ matches the relative entropy $D(p^*\|p)$ between bivariate normal densities,
where $p(z,\tilde z)$ is the joint density of $Z,\tilde Z$ of correlation $\rho$ and where $p^*(z,\tilde z)$ is the joint normal obtained by tilting that density by $e^{(\lambda/2)(z^2-\tilde z^2)}$, chosen to make $(1/2)(Z^2-\tilde Z^2)$ have mean $\Delta$, when there is such a $\lambda$.

Let's give $D(\Delta,1\!-\!\rho^2)$ explicitly as an increasing function of the ratio $\Delta^2/(1\!-\!\rho^2)$.  Working with logarithm base $e$,
the derivative with respect to $\lambda$ of the expression being maximized yields a quadratic equation which can be solved for the optimal
$$\lambda^*= \frac{1}{2\Delta}\big( \sqrt {1 + 4\Delta^2/(1 \! - \! \rho^2)} - 1 \big).$$
Let $q = 4\Delta^2/(1 \!-\! \rho^2)$ and $\gamma=\sqrt{1+q}-1$, which is near $q/2$ when $q$ is small and approximately $\sqrt q$ when $q$ is large. Plug the optimized $\lambda$ into the above expression and simplify
to obtain
$D= (1/2)\big(\gamma - \log(1\!+\!\gamma/2)\big)$,
which is at least $\gamma/4$.
Thus $D$ is the composition of strictly increasing non-negative functions $(1/2)\big(\gamma - \log(1+\gamma/2)\big)$ and $\gamma=\sqrt{1\!+\!q}-1$ evaluated at $q= 4\Delta^2/(1\!-\!\rho^2)$.
For small values of this ratio, we see that $D$ is near $q/8=
(1/2){\Delta^2}/(1 \! - \! \rho^2)$.

The expression corresponding to $D$ 
but with the maximum restricted to $0 \!\le\! \lambda \!\le\! 1$ is denoted $D_1\!=\!D_1(\Delta,1\!-\!\rho^2)$, that is,
$$D_1= \max_{0\le \lambda \le 1} \big\{ \lambda \Delta + (1/2) \log (1- \lambda^2(1 \!-\! \rho^2))\big\}.$$
The corresponding optimal value of $\lambda$ is $\min\{1,\lambda^*\}$.
When the optimal $\lambda$ is less than $1$, the value of $D_1$ matches $D$ as given above.  

The $\lambda \!=\! 1$ case occurs when $1+ 4 \Delta^2/(1\!-\!\rho^2) \ge (1+ 2\Delta)^2$, or equivalently $\Delta \ge (1\!-\!\rho^2)/\rho^2$.  Then
the exponent is $D_1= \Delta + (1/2) \log \rho^2$,
which is as least $\Delta - (1/2)\log (1+\Delta)$. Consequently, in this regime $D_1$ is between $\Delta/2$ and $\Delta$.

The special case $\rho^2=1$ is included with $D_1=\Delta$.


\section{Performance of Least Squares}
\label{sec:rells}
As we have said, least squares provides optimal decoding of superposition codes.
In this section we examine the performance of this least squares choice in terms of rate and reliability.  We focus on partitioned superposition codes in which the codewords are superpositions with one term from each section.

Let $S$ be an allowed subset of terms.  We examine first subset coding in which to each such $S$ there is a corresponding coefficient vector $\beta$ in which the non-zero coefficients take a specified positive value as discussed above. We may denote the corresponding codeword $X_S=X\beta$. Among such codewords, least squares provides a choice for which $|Y-X_{S}|^2$ is minimal.


For a subset $S$ of size $L$ we measure how different it is from $S^*$, the subset that was sent.  Let $\ell=card(S-S^*)$ be the number of entries of $S$ not in $S^*$. Equivalently, since $S$ and $S^*$ are of the same size, it is the number of entries of $S^*$ not in $S$.

Let $\hat S$ be the least squares solution, or an approximate least squares solution, achieving $|Y-X_{\hat S}|^2 \le |Y-X_{S^*}|^2 +\delta_0$ with $\delta_0 \ge 0$.
We call $card(\hat S- S^*)$ the number of mistakes.  Indeed, for a partitioned superposition code it is the number of sections incorrectly decoded.


There is a role for
the function $C_\alpha = \frac{1}{2} \log(1+\alpha v)$ for $0 \le \alpha \le 1$, where $v= P/\sigma^2$ is the signal-to-noise ratio and $C_1=C=(1/2)\log(1+v)$ is the channel capacity.  We note that $C_\alpha - \alpha C$ is a non-negative concave function equal to $0$ when $\alpha$ is $0$ or $1$ and strictly positive in between.
The quantity $C_\alpha - \alpha R$ is larger by the additional amount $\alpha (C-R)$, positive when the rate $R$ is less than the Shannon capacity $C$.

The function $\psi_\alpha (\lambda)= -(1/2)\log[1-\lambda^2 \alpha v/(1+\alpha v)]$
with $0 \!\le\! \lambda \!\le\! 1$ is 
the cumulant generating function of a test statistic in our analysis.



Our first result on the distribution of the number of mistakes is the following. 

\noindent
{\bf Lemma 1:} Set $\alpha = \ell/L$ for an $\ell \in \{1,2,\ldots,L\}$. For  approximate least squares
with $0 \le \delta_0 \le 2 \sigma^2 (C_\alpha -\alpha R)/\log e$, the probability of a fraction $\alpha=\ell/L$ mistakes is upper bounded by
$${L \choose {\alpha L}} \exp\left\{-n \max_{0\le \lambda \le 1} \big\{\lambda \Delta_\alpha - \psi_\alpha (\lambda)\big\}\right\},$$
or equivalently,
$${L \choose {\alpha L}} \exp\left\{-n D_1( \Delta_\alpha,\alpha v/(1+\alpha v))\right\},$$
where $\Delta_\alpha = C_\alpha - \alpha R - (\delta_0/2\sigma^2)\log e$ and $v$ is the signal-to-noise ratio.

\vspace{.3cm}
\noindent{\bf Remark 1:} We find this Lemma 1 to be especially useful for $\alpha$
in the lower range of the interval from $0$ to $1$.  
Lemma 2 below will refine the analysis to provide an
exponent more useful in the upper range of the interval.



\vspace{.3cm}
\noindent
{\bf Proof of Lemma 1:}
To incur $\ell$ mistakes, there must be an allowed subset $S$ of size $L$ which differs from the subset $S^*$ sent in an amount $card(S-S^*)=card(S^*-S)=\ell$
which undesirably has squared distance $|Y-X_S|^2$ less than or equal to the value $|Y-X_{S^*}|^2+\delta_0$ achieved by $S^*$.

The analysis proceeds by considering an arbitrary such $S$, bounding the probability that $|Y-X_{S}|^2 \le |Y-X_{S^*}|^2 +\delta_0$, and then using an appropriately designed union bound to put such probabilities together.

Consider the statistic $T=T(S)$ given by
$$ T(S)= \frac{1}{2} \left[ \frac{|Y-X_S|^2}{\sigma^2} \, - \, \frac{|Y-X_{S^*}|^2}{\sigma^2} \right].$$
We set a threshold for this statistic equal to $t = \delta_0/(2\sigma^2)$.
The event of interest is that $T \le t$.

The subsets $S$ and $S^*$ have an intersection $S_{1}= S \intersect S^*$ of size $L - \ell$ and difference $S_2 = S-S_1$ of size $\ell=\alpha L$.  Given $(\X_j: j \in S)$ the actual density of $Y$ is normal with mean $X_{S_{1}} = \sum_{j \in S_{1}} X_j$ and variance $(\sigma^2 + \alpha P) I$ and we denote this density $p(Y|X_{S_1})$.  In particular, there is conditional independence of $Y$ and $X_{S_2}$ given $X_{S_1}$.

Consider the alternative hypothesis of a conditional distribution for $Y$ given $X_{S_{1}}$ and $X_{S_2}$ which is Normal($X_S,\sigma^2 I$). It is the distribution which would have governed $Y$ if $S$ were sent. Let $p_h(Y|X_{S_1},X_{S_2})=p_h(Y|X_S)$ be the associated conditional density.
With respect to this alternative hypothesis, the conditional distribution for $Y$ given $X_{S_1}$ remains Normal($X_{S_1},(\sigma^2+\alpha P)I$).
That is, $p_h(Y|X_{S_1})=p(Y|X_{S_1})$.

We decompose the above test statistic as
$$ \frac{1}{2} \left[ \frac{|Y-X_{S_{1}}|^2}{\sigma^2 +\alpha P} - \frac{|Y-X_{S^*}|^2}{\sigma^2} \right]$$
$$+ \, \frac{1}{2} \left[ \frac{|Y-X_S|^2}{\sigma^2} - \frac{|Y-X_{S_1}|^2}{\sigma^2 +\alpha P} \right].$$
Let's call the two parts of this decomposition $T_1$ and $T_2$, respectively.  Note that $T_1=T_1(S_1)$ depends only on terms in $S^*$, whereas $T_2=T_2(S)$ depends also on the part of $S$ not in $S^*$.


Concerning $T_2$, note that we may express it as
$$T_2(S) = \frac{1}{n} \log \frac {p(Y|X_{S_1})}{p_h(Y|X_{S})} + C_{\alpha},$$
where
$$C_{\alpha} = \frac{1}{2} \log \frac{\sigma^2 + \alpha P}{\sigma^2}$$
is the adjustment by the logarithm of the ratio of the normalizing constants of these densities.

Thus $T_2$ is equivalent to a likelihood ratio test statistic between the actual conditional density and the constructed alternative hypothesis for the conditional density of $Y$ given $X_{S_1}$ and $X_{S_2}$.  
It is helpful to use Bayes rule 
to provide $p_h(X_{S_2}|Y,X_{S_1})$ via the equality of $\frac{p_h(X_{S_2}|Y,X_{S_1})}{p(X_{S_2}|X_{S_1})}$ and $\frac{p_h(Y|X_{S_1},X_{S_2})}{p(Y|X_{S_1})}$ and to interpret this equality as providing an alternative representation of the likelihood ratio in terms of the reverse conditionals for $X_{S_2}$ given $X_{S_1}$ and $Y$.


We are examining the event $E_\ell$ that there is an allowed subset $S=S_1\union S_2$ (with $S_1=S \intersect S^*$ of size $L-\ell$ and $S_2=S-S_1$ of size $\ell$) such that that $T(S)$ is less than $t$. For positive $\lambda$ the indicator of this event satisfies

$$1_{E_\ell} \le \sum_{S_1} \left( \sum_{S_2}  e^{-n(T(S)-t)} \right)^\lambda,$$
because, if there is such an $S$ with $T(S)-t$ negative, then indeed that contributes a term on the right side of value at least $1$.  Here the outer sum is over $S_1 \subset S^*$ of size $L-\ell$.  For each such $S_1$, for the inner sum, we have $\ell$ sections in each of which, to comprise $S_2$, there is a term selected from among $B-1$ choices other than the one prescribed by $S^*$.

To bound the probability of $E_\ell$, take the expectation of both sides, bring the expectation on the right inside the outer sum, and write it as the iterated expectation, where on the inside condition on $Y$, $X_{S_1}$ and $X_{S^*}$ to pull out the factor involving $T_1$, to obtain that $\PP[E_\ell]$ is not more than
$$\sum_{S_1} 
\E e^{-n\lambda(T_1(S_1)-t)} \E_{X_{S_2}|Y,X_{S_1},X_{S^*}} \left(\sum_{S_2} e^{-nT_2(S)} \right)^\lambda .$$
A simplification here is that the true density for $X_{S_2}$ is independent of the conditioning variables $Y$, $X_{S_1}$ and $X_{S^*}$.

We arrange for $\lambda$ to be not more than $1$.  Then 
by Jensen's inequality, the conditional expectation may be brought inside the $\lambda$ power and inside the inner sum, yielding
$$\PP[E_\ell] \le \sum_{S_1} 
\E e^{-n\lambda(T_1(S_1)-t)} \left(\sum_{S_2} \E_{X_{S_2}|Y,X_{S_1}} e^{-nT_2(S)} \right)^\lambda .$$
Recall that $$e^{-nT_2(S)} = \frac{p_h(X_{S_2}|Y,X_{S_1})}{p(X_{S_2})} e^{-nC_\alpha}$$ and that the true density for $X_{S_2}$ 
is independent of the conditioning variables in accordance with the $p(X_{S_2})$ in denominator.  So when we take the expectation of this ratio we cancel the denominator leaving the numerator density which integrates to $1$.  Consequently, the resulting expectation of $e^{-nT_2(S)}$ is not more than $e^{-nC_\alpha}$.  The sum over $S_2$ entails less than $B^\ell= e^{nR\ell/L}$ choices so the bound is
$$\PP[E_\ell] \le \sum_{S_1} \E e^{-n\lambda T_1(S_1)} e^{-n\lambda[C_\alpha - \alpha R-t]}.$$
Now $n T_1(S_1)$ is a sum of $n$ independent mean-zero random variables each of which is the difference of squares of normals for which the squared correlation is 
$\rho_\alpha^2=1/(1\!+\! \alpha v)$.
So the expectation $\E e^{-n\lambda T_1(S_1)}$ is found to be equal to $[1/[1-\lambda^2 \alpha v/(1+\alpha v)]]^{n/2}$.
When plugged in above it yields the claimed bound optimized over $\lambda$ in $[0,1]$.  We recognize that the exponent takes the form $D_1(\Delta,1\!-\!\rho^2)$
with $1\!-\!\rho^2= \alpha v/(1\!+\!\alpha v)$ as discussed in the preliminaries.  This completes the proof of Lemma 1.

\vspace{.3cm}
\noindent{\bf Some additional remarks:}
The exponent $D_1$ in Lemma 1 (and its refinement in Lemma 2 to follow)
depends on the fraction of mistakes $\alpha$ and the signal-to-noise ratio $v$ only through $\Delta_\alpha = C_\alpha - \alpha R - t$ and $1\!-\!\rho_\alpha^2$.
As we have seen, the $\lambda \!<\! 1$ case occurs when $\Delta_\alpha <  (1\!-\!\rho_\alpha^2)/\rho_\alpha^2$
and then 
$D$ is near $(1/2) \Delta_\alpha^2/(1 \! - \! \rho_\alpha^2)$ when it is small; whereas, the $\lambda \!=\! 1$ case occurs
when $\Delta_\alpha \ge (1\!-\!\rho_\alpha^2)/\rho_\alpha^2$ and then
the exponent
is as least $\Delta_\alpha - (1/2)\log (1+\Delta_\alpha) \ge \Delta_\alpha/2$. 

This behavior of the exponent is similar to the usual order $(C \!-\! R)^2$
for $R$ close to $C$ and order $C\!-\!R$ 
for $R$ farther from $C$ associated with the theory in Gallager \cite{Gallager1968}.

A difficulty with the Lemma 1 bound is that for $\alpha$ near $1$ and for $R$ correspondingly close to $C$, in the key quantity $\Delta_\alpha^2/(1\!-\!\rho_\alpha^2)$, the order of $\Delta_\alpha^2$ is $(1\!-\!\alpha)^2$, which is too close to zero to cancel the effect of the combinatorial coefficient.

The following lemma refines the analysis of Lemma 1, obtaining the same exponent with an improved correlation coefficient. The denominator  $1\!-\!\rho_\alpha^2 = \alpha(1\!-\!\alpha)/(1\!+\!\alpha v)$ is improved by the presence of the factor $(1\!-\!\alpha)$ allowing the conclusion to be useful also for $\alpha$ near $1$.
The price we pay is 
the presence of an additional term in the bound.


For the statement of Lemma 2 we again use the test statistic $T(S)$ as defined in the proof of Lemma 1.  For interpretation of what follows with arbitrary base of logarithm, in that definition of $T(S)$ multiply by $\log e$ and likewise take the threshold to be $t=\frac{\delta_0}{2\sigma^2} \log e$.


\vspace{0.5cm}
\noindent
{\bf Lemma 2:}  Let a positive integer $\ell \le L$ be given and let $\alpha=\ell/L$.  Suppose $0 \le t < C_\alpha - \alpha R$. As above let $E_\ell$ be the event that there is an allowed $L$ term subset $S$ with $S-S^*$ of size $\ell$ such that $T(S)$ is less than $t$.
Then $\PP[E_\ell]$ is bounded by the minimum for $t_\alpha$ in the interval between $t$ and $C_\alpha - \alpha R$ of the following

$$\binom{L}{L\alpha}\exp\left\{-nD_1(\calph - \alpha R - t_\alpha,1\!-\!\rho_\alpha^2)\right\} $$
$$+ \exp\big\{-nD(t_\alpha \!-\! t,\alpha^2v/(1\!+\!\alpha^2 v)]\big\}.
$$
where
$1\!-\!\rho_\alpha^2 = \alpha(1\!-\!\alpha) v/(1+\alpha v).$




\vspace{.3cm}
\noindent
{\bf Proof of Lemma 2:} Split the test statistic $T(S)=\tilde T(S)+T^*$ where
$${\tts = \frac{1}{2} \left[ \ymxs - \ymaxst \right]}$$
and
$${T^* = \frac{1}{2} \left[ \ymaxst - \frac{|Y-X_{S^*}|^2}{\sigma^2} \right]}$$
Likewise we split the threshold $t= \tilde t+ t^*$ where $t^*= -(t_\alpha-t)$ is negative and $\tilde t= t_\alpha$ is positive.

The event that there is an $S$ with $T(S) \!<\! t$ is contained in the union of the two events $\tilde E_\ell$, that there is an $S$ with $\tilde T(S) \!<\! \tilde t$, and the event $E_\ell^*$, that $T^* \!<\! t^*$.
The part $T^*$ has no dependence on $S$ so it can be treated more simply. It is a mean zero average of differences of squared normal random variables, with squared correlation $1/(1+\alpha^2 v)$.  So using its moment generating function, $\PP[E_\ell^*]$ is exponentially small, bounded by the second of the two expressions above.

Concerning $\PP[{\tilde E}_\ell]$, its analysis is much the same as for Lemma 1.  We again decompose $\tilde T(S)$ as the sum $\tilde T_1 (S_1)+ \tilde T_2(S)$, where $\tilde T_2(S)=T_2(S)$ is the same as before. The difference is that in forming $\tilde T_1 (S_1)$ we subtract $\ymaxst$ rather than $\frac{|Y-X_{S^*}|^2}{\sigma^2}$.  
Consequently,
$$\tilde T_1 (S_1)= \frac{1}{2}\left[\frac{|Y-X_{S_1}|^2}{\sigma^2 + \alpha P} - \ymaxst \right],$$
which again involves a difference of squares of standardized normals. But here the coefficient $(1\!-\!\alpha)$ multiplying $X_{S^*}$ is such that we have maximized the correlations between the $Y-X_{S_1}$ and $Y-(1\!-\!\alpha)X_{S^*}$. Consequently, we have reduced the spread of the distribution of the differences of squares of their standardizations as quantified by the cumulant generating function.  One finds that the squared correlation coefficient is $\rho_\alpha^2 = (1+\alpha^2 v)/(1+\alpha v)$ for which $1\!-\!\rho_\alpha^2 = \alpha(1\!-\!\alpha)v/(1+\alpha v)$. Accordingly we have that the moment generating function is $E e^{- n\lambda \tilde T(S_1)}= \exp\{-(n/2)\log[1-\lambda^2(1\!-\!\rho_\alpha^2)]\}$ which gives rise to the bound appearing as the first of the two expressions above.
This completes the proof of Lemma 2.


\begin{figure}
\centerline{\includegraphics[height=3.5in]{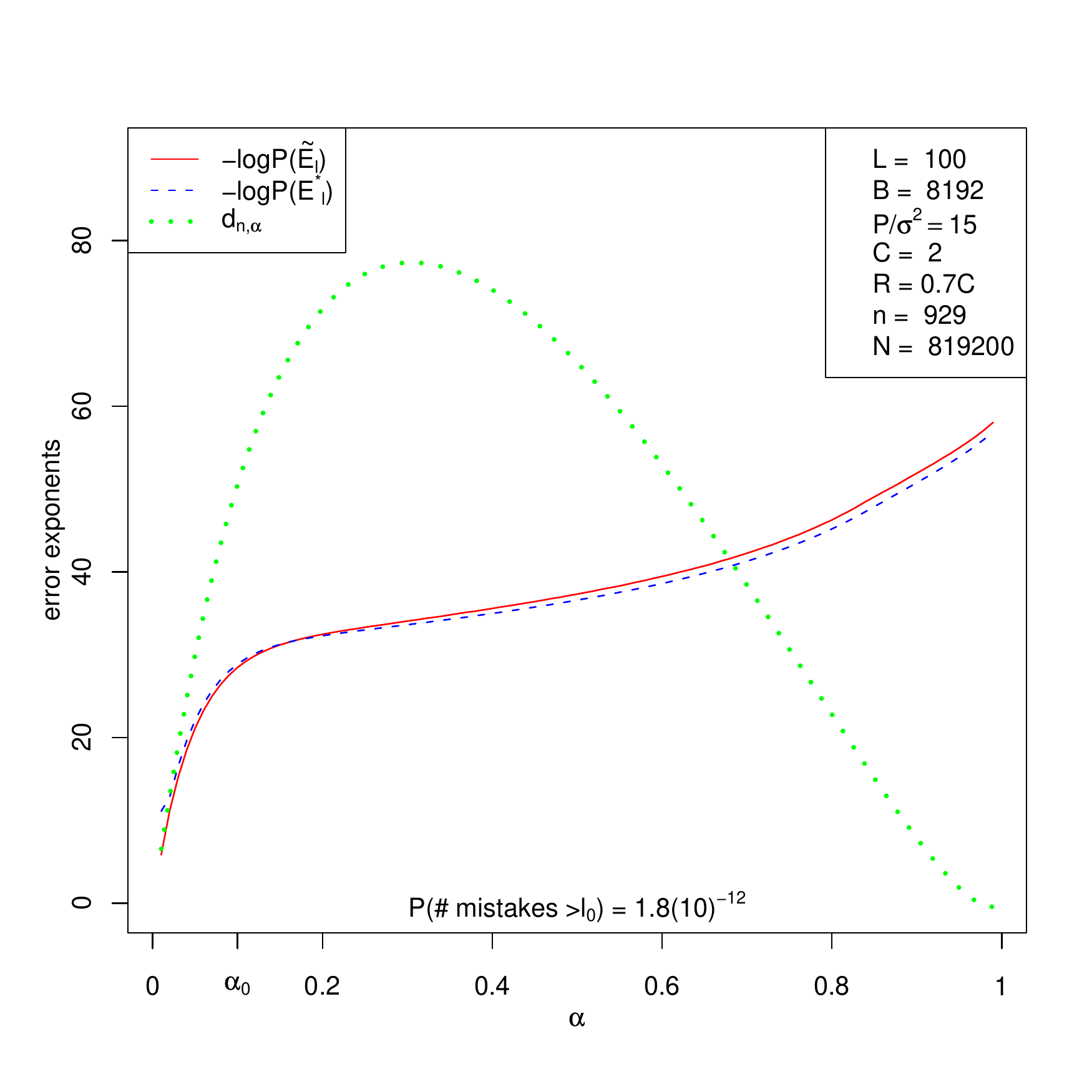}}
\caption{Exponents of contributions to the error probability as functions of $\alpha=\ell/L$ using exact least squares, i.e., $t=0$, with $L=100$,$\,B=2^{13}$, signal-to-noise ratio $v=15$, and rate $70\%$ of capacity. The red and blue curves are the $-\log \PP[\tilde E_\ell]$ and $-\log \PP[E_\ell^*]$ bounds, using the natural logarithm, from the two terms in Lemma 2 with optimized $t_\alpha$.  The dotted green curve is $d_{n,\alpha}$ explained below. With $\alpha_0 = 0.1$, the total probability of at least that fraction of mistakes is bounded by $1.8(10)^{-12}$.}
\label{fig1}
\end{figure}

\vspace{0.2cm}

The method of analysis also allows consideration of subset coding 
without partitioning. For, in this case all $N \choose L$ subsets of size $L$ correspond to codewords, so with the rate in nats we have $e^{nR} = {N \choose L}$.  The analysis proceeds in the same manner, with the same number $L \choose {L-\ell}$ of choices of sets $S_1=S\intersect S^*$ where $S$ and $S^*$ agree on $L-\ell$ terms, but now with ${N\!-\!L} \choose \ell$ choices of sets $S_2 = S\!-\!S^*$ of size $\ell$ where they disagree.  We obtain the same bounds as above except that where we have $B^\ell = e^{n \alpha R}$ with the exponent $\alpha R$ it is replaced by ${{N\!-\!L} \choose \ell} = e^{nR(\alpha)}$ with the exponent $R(\alpha)$ defined by 
$R(\alpha)= R \log {{N\!-\!L} \choose \alpha L}/\log {N \choose L}$. Thus we have the following conclusion.

{\bf Corollary 3:}
For subset superposition coding, the probability of the event $E_{\ell}$ that there is a $\beta$ that is incorrect in $\ell$ sections and has $|Y-X\beta|^2 \le |Y-X\beta^*|^2 + \delta_0$ is bounded by the minimum of the same expressions given in Lemma 1 and Lemma 2 except that the term $\alpha R$ appearing in these expression be replaced by the quantity $R(\alpha)$ defined above.

\vspace{.3cm}
\section{Sufficient Section Size}
\label{sec:sufsec}
We come to the matter of sufficient conditions on the section size $B$ for our exponential bounds to swamp the combinatorial coefficient, for partitioned superposition codes.

We call $a= (\log B)/(\log L)$ the \emph{section size rate}, that is, the bits required to describe the member of a section relative to the bits required to describe which section.  It is invariant to the base of the log.  Equivalently we have $B$ and $L$ related by $B=L^a$.  Note that the size of $a$ controls the polynomial size of the dictionary $N=BL=L^{a+1}$.


In both cases the codelength may be written as $$n = \frac{a L \log L}{R}.$$

We do not want a requirement on the section sizes with $a$ of order $1/(C\!-\!R)$ for then the complexity would grow exponentially with this inverse of the gap from capacity.  So instead let's decompose $\dalp = \tdalp + \alpha (C\!-\!R) - t_\alpha$ where $\tdalp = C_\alpha - \alpha C$. We investigate in this section the use of $\tdalp$ to swamp the combinatorial coefficient. In the next section excess in $\tdalp$, beyond that needed to cancel the combinatorial coefficient, plus $\alpha (C\!-\!R) - t_\alpha$ are used to produce exponentially small error probability.

Define $D_{\alpha,v}\!=\!D_1(\dalp,1\!-\!\rho_\alpha^2)$ and ${\tilde D}_{\alpha,v}\!=\!D_1({\tilde \Delta}_\alpha,1\!-\!\rho_\alpha^2)$.
Now $D_1(\Delta,1\!-\!\rho^2)$ is increasing as a function of $\Delta$, so $D_{\alpha,v}$ is greater than ${\tilde D}_{\alpha,v}$ whenever $\dalp > \tdalp$.  Accordingly, we decompose the exponent $D_{\alpha,v}$ as the sum of two components, namely, ${\tilde D}_{\alpha,v}$ and the difference $D_{\alpha,v}-{\tilde D}_{\alpha,v}$.

We then ask whether the first part of the exponent denoted ${\tilde D}_{\alpha,v}$
is sufficient to wash out the affect of the log combinatorial coefficient $\log \binom {L} {L \alpha}$.  That is, we want to arrange for the nonnegativity of
the difference $$d_{n,\alpha} = n {\tilde D}_{\alpha,v}
- \log \binom {L} {L \alpha}.$$
This difference is small for $\alpha$ near $0$ and $1$.
Furthermore, its constituent quantities have a shape comparable to multiples of $\alpha(1\!-\!\alpha)$.
Consider first $\tdalp= C_\alpha - \alpha C$ and take the log to be base $e$.  It has second derivative $-(1/2)v^2/(1+\alpha v)^2$.
It follows that $\tdalp \ge (1/4)\alpha(1\!-\!\alpha) v^2/ (1+v)^2$, since the difference of the two sides has negative second derivative, so it is concave and equals $0$ at $\alpha\!=\!0$ and $\alpha\!=\!1$.  Likewise $(1\!-\!\rho_\alpha^2) = \alpha(1\!-\!\alpha)v/(1\!+\!\alpha v)$ so the ratio $\tilde u = 4 \tdalp^2/(1\!-\!\rho_\alpha^2)$ is at least $(1/4)\alpha(1\!-\!\alpha)v^3(1\!+\!\alpha v)/(1+v)^4$.  Consequently, whether the optimal $\lambda$ is equal to $1$ or is less than $1$, we find that ${\tilde D}_{\alpha,v}$ is of order $\alpha(1\!-\!\alpha)$.

Similarly, there is the matter of $\log \binom {L}{L\alpha}$, with $L \alpha$ restricted to have integer values. It enjoys the upper bounds $\min (\alpha,1\!-\!\alpha) L\log L$ and $L \log 2$ so that it is not more than $\alpha(1\!-\!\alpha) (L \log L)/(1\!-\! \delta_L)$ where $\delta_L = (\log 2)/\log L$.

Consequently, using $n=(a L \log L)/R$, one finds that for sufficiently large $a$ depending on $v$, 
the difference $d_{n,\alpha}$ is nonnegative uniformly for the permitted $\alpha$ in $[0,1]$. The smallest such section size rate is
$$a_{v,L}= \max_\alpha \frac{R \log \binom {L} {L\alpha}}{ {\tilde D}_{\alpha,v} \, L \log L},$$
where the maximum is for $\alpha$ in $\{1/L,2/L,\ldots,1-1/L\}$.
This definition has the required invariance to the choice of base of the logarithm, assuming that the same base is used for the communication rate $R$ and for the $C_\alpha-\alpha C$ that arises in the definition of ${\tilde D}_{\alpha,v}$.

In the above ratio the numerator and denominator are both $0$ at $\alpha\!=\!0$ and $\alpha\!=\!1$ (yielding $d_{n,\alpha} \!=\!0$ at the ends). Accordingly, we have excluded $0$ and $1$ from the definition of $a_{v,L}$ for finite $L$.  Nevertheless, limiting ratios arise at these ends.

We show that the value of $a_{v,L}$ is fairly insensitive to the value of $L$, with the maximum over the whole range being close to a limit $a_v$ which is characterized by values in the vicinity of $\alpha=1$.

Let $v^*$ near $15.8$ be the solution to $(1\!+\!v^*)\log(1\!+\!v^*)=3v^*\log e.$ 

\vspace{0.3cm}
\noindent
{\bf Lemma 4:} The section size rate $a_{v,L}$ has a continuous limit $a_v=\lim_{L\rightarrow \infty } a_{v,L}$ which is given, for $0 < v < v^*$, by
$$a_v=\frac{R}{[(1\!+\!v)\log(1\!+\!v) - v\log e]^2 /[8v(1\!+\!v)\log e]}$$
and for $v \ge v^*$ by
$$a_v=\frac{R}{[(1\!+\!v)\log(1\!+\!v) - 2v\log e]/[2(1\!+\!v)]}$$
where $v$ is the signal-to-noise ratio.
With $R$ replaced by 
$C=(1/2)\log(1\!+\!v)$ and using log base e, in the case $0 \!<\! v \!<\! v^*$, it is
$$\frac{4v(1\!+\!v)\log(1\!+\!v)}{[(1\!+\!v)\log(1\!+\!v) - v]^2 }$$
which is approximately $16/v^2$ for small positive $v$; whereas, in the case $v \ge v^*$
it is
$$\frac{(1\!+\!v)\log(1\!+\!v)}{(1\!+\!v)\log(1\!+\!v)- 2v}$$
which asymptotes to the value $1$ for large $v$.


\vspace{0.3cm}
\noindent
{\bf Proof of Lemma 4:} For $\alpha$ in $(0,1)$ we use $\log \binom {L} {L\alpha} \le L \log 2$ and the strict positivity of ${\tilde D}_{\alpha,v}$ to see that the ratio in the definition of $a_{v,L}$ tends to zero 
uniformly within compact sets interior to $(0,1)$.  So the limit $a_v$ is determined by the maximum of the limits of the ratios at the two ends.  In the vicinity of the left and right ends we replace $\log \binom {L}{L\alpha}$ by the continuous upper bounds
$\alpha L \log L$ and $(1\!-\!\alpha) L \log L$, respectively, which are tight at $\alpha = 1/L$ and $1\!-\!\alpha=1/L$, respectively.  Then in accordance with L'Hopital's rule, the limit of the ratios equals the ratios of the derivatives at $\alpha \! =\! 0$ and $\alpha \!=\! 1$, respectively.  Accordingly,
$$a_v = \max \left\{ \frac{R}{{\tilde D}_{0,v}^\prime}, \frac{-R}{{\tilde D}_{1,v}^\prime}\right\},$$
where ${\tilde D}_{0,v}^\prime$ and ${\tilde D}_{1,v}^\prime$ are the derivatives of ${\tilde D}_{\alpha,v}$ with respect to $\alpha$ evaluated at $\alpha\!=\!0$ and $\alpha\!=\!1$, respectively.

To determine the behavior of ${\tilde D}_\alpha= {\tilde D}_{\alpha,v}$ in the vicinity of $0$ and $1$
we first need to determine whether the optimal $\lambda$ in its definition is strictly less than $1$ or equal to $1$.  According to our earlier developments that is determined by whether ${\tilde \Delta}_\alpha < (1\!-\!\rho_\alpha^2)/\rho_\alpha^2$.
The right side of this is $\alpha(1\!-\!\alpha)v/(1+\alpha^2 v)$.  So it is equivalent to determine whether the ratio
$$\frac{(C_\alpha - \alpha C)(1+\alpha^2 v)}{\alpha(1\!-\!\alpha)v}$$
is less than $1$ for $\alpha$ in the vicinity of $0$ and $1$.  Using L'Hopital's  rule 
it suffices to determine whether the ratio of derivatives
is less than $1$ when 
evaluated at $0$ and $1$.  At $\alpha=0$ it is $(1/2)[v- \log(1+v)]/v$ which is not more than $1/2$ (certainly less than $1$) for all positive $v$; whereas,
at $\alpha=1$ the ratio of derivatives is $(1/2)[(1+v)\log (1+v) - v]/v$ which is less than $1$ if and only if $v < v^*$.

For the cases in which the optimal $\lambda < 1$, we need to determine the derivative of ${\tilde D}_\alpha$ at $\alpha\!=\!0$ and $\alpha\!=\!1$.  Recall that $\tilde D_\alpha$ is the composition of the functions $(1/2)(\gamma- \log(1+\gamma/2))$ and $\gamma= \sqrt{1+u}-1$ and
$u_\alpha= 4{\tilde \Delta}_\alpha^2 /(1\!-\!\rho_\alpha^2)$.  We use the chain rule taking the products of the associated derivatives.  The first of these functions has derivative $(1/2)(1- 1/(2+\gamma))$ which is $1/4$ at $\gamma\!=\!0$, the second of these has derivative $1/(2\sqrt{1\!+\!u})$ which
is $1/2$ at $u\!=\!0$, and the third of these functions is
$$u_\alpha= \frac{\big(\log(1+\alpha v) - \alpha \log(1\!+\!v)\big)^2}{\alpha(1\!-\!\alpha)v/(1+\alpha v)}$$
which has derivative 
that evaluates to
${(v-\log(1\!+\!v))^2}/v$ at $\alpha \!=\! 0$ and evaluates to $- [(1\!+\!v)\log(1\!+\!v)-v]^2/[v(1\!+\!v)]$ at $\alpha\!=\!1$.  The first of these gives what is needed for the left end for all positive $v$ and the second what is needed for the right end for all $v < v^*$.

The magnitude of the derivative at $1$ is smaller than at $0$.
Indeed, taking square roots this is the same as the claim that
$(1\!+\!v)\log(1\!+\!v)-v < \sqrt{1\!+\!v} (v-\log(1\!+\!v))$.
Replacing $s=\sqrt{1\!+\!v}$ and rearranging,
it reduces to $s \log s < (s^2-\!1)/2$, which is 
true for $s \!>\! 1$ since the two sides match at $s=1$ and have derivatives $1+\log s < s$. Thus the limiting value for $\alpha$ near $1$ is what matters for the maximum.
This produces the claimed form of $a_v$ for $v < v^*$.

In contrast for $v > v^*$, the optimal $\lambda \!=\! 1$ for $\alpha$ in the vicinity of $1$.  In this case we use ${\tilde D}_\alpha= \tilde \Delta_\alpha + (1/2)\log \rho_\alpha^2$ which has derivative 
equal to
$-(1/2) [(1\!+\!v)\log(1\!+\!v) - 2v]/(1\!+\!v)$ at $\alpha\!=\!1$, which is again smaller in magnitude than the derivative at $\alpha\!=\!0$, producing the claimed form for $a_v$ for $v > v^*$.

At $v=v^*$ we equate $(1+v)\log(1+v)=3v$ and see that both of the expressions for the magnitude of the derivative at $1$ agree with each other (both reducing to $v/(2(1+v))$) so the argument extends to this case, and the expression for $a_v$ is continuous in $v$.  This completes the proof of Lemma 3.

While $a_v$ is undesirably large for small $v$, we have reasonable values for moderately large $v$. In particular, $a_v$ equals $5.0$ and $3$, respectively, 
at $v=7$ and $v^*=15.8$, and it is near $1$ for large $v$.



\begin{figure}
\centerline{\includegraphics[height=3.5in]{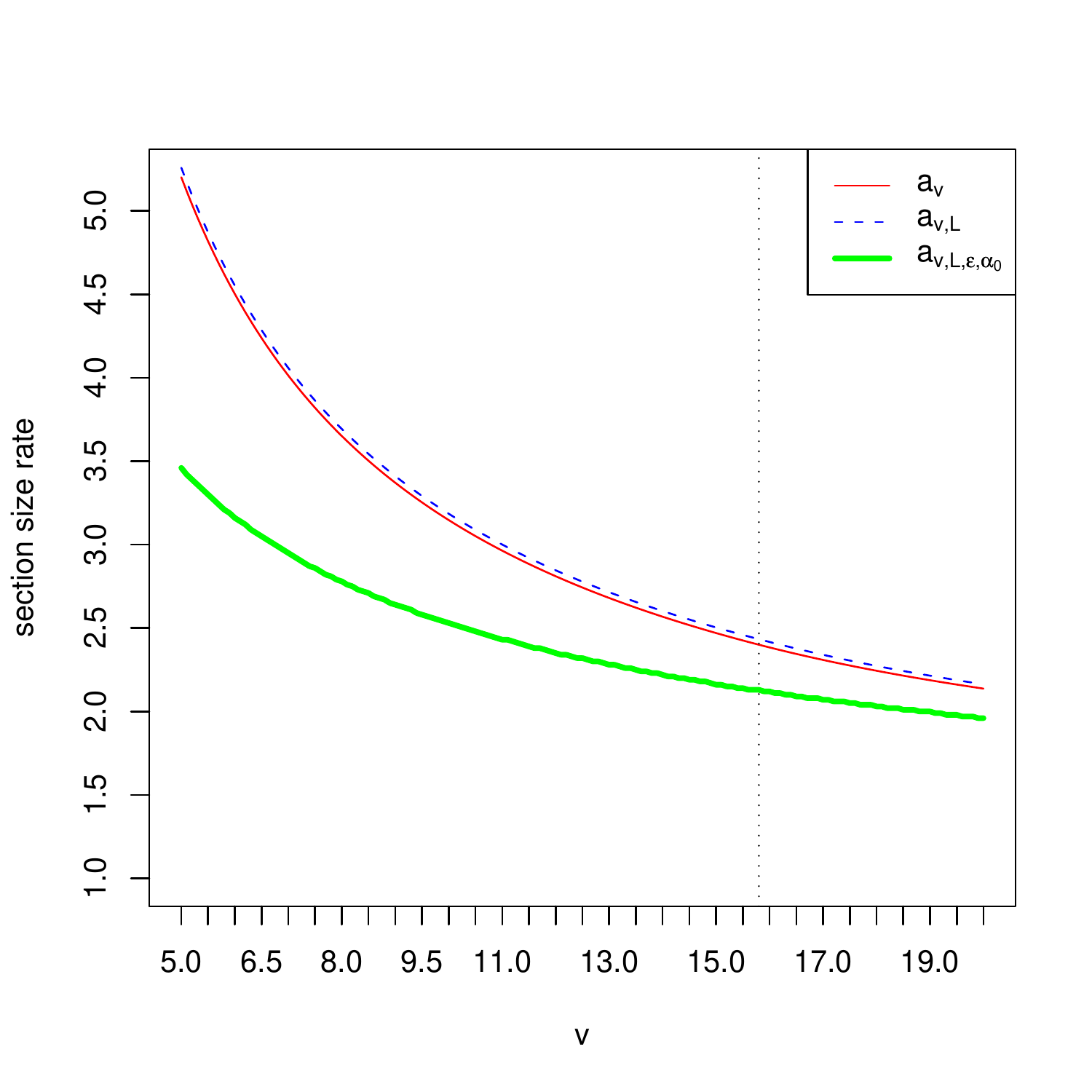}}
\caption{Sufficient section size rate $a$ as a function of the signal-to-noise ratio $v$. The dashed curve shows $a_{v,L}$ at $L=64$. Just below it the thin solid curve is the limit for large $L$. For section size $B \ge L^{a}$ the error probabilities are exponentially small for all $R < C$ and any $\alpha_0> 0$. The bottom curve shows the minimal section size rate for the bound on the error probability contributions to be less than $e^{-10}$, with $R=0.8C$ and $\alpha_0=0.1$ at $L=64$.
}
\label{fig:secsizerate}
\end{figure}

Numerically is of interest to ascertain the minimal section size rate $a_{v,L,\epsilon,\alpha_0}$, for a specified $L$ such as $L=64$, for $R$ chosen to be a proscribed high fraction of $C$, say $R=0.8 C$, for $\alpha_0$ a proscribed small target fraction of mistakes, say $\alpha_0=0.1$, and for $\epsilon$ to be a small target probability, so as to obtain $\min\{P[E_\ell],P[\tilde E_\ell]+P[E_\ell^*]\}\le \epsilon$, taking the minimum over allowed values of $t_\alpha$, for every $\alpha = \ell/L$ at least $\alpha_0$. For this calculation the bound from Lemma 1 is used for $P[E_\ell]$ and the bound from Lemma 2 is used for $P[\tilde E_\ell]+P[E_\ell^*]$. This is illustrated in Figure \ref{fig:secsizerate} plotting the minimal section size rate as a function of $v$ for $\epsilon = e^{-10}$.  With such $R$ moderately less than $C$ we observe substantial reduction in the required section size rate.


\vspace{0.3cm}
\noindent
{\bf Extra $\Delta_\alpha$ beyond the minimum:}  Via the above analysis we determine the minimum value of $\Delta$ for which the combinatorial term is canceled, and we characterize the amount beyond that minimum which makes the error probability exponentially small.  
Arrange $\Delta_\alpha^{\min}$ to be the solution to the equation $$nD_1(\Delta_\alpha^{\min},1-\rho_\alpha^2) = \log  {L \choose {L\alpha} }.$$  To see its characteristics,
let $\Delta_\alpha^{target} = (1-\rho_\alpha^2)^{1/2}G(r_\alpha)$ at $$r_\alpha =\frac{1}{n} \log  {L \choose {L\alpha} },$$
using log base $e$. Here $G(r)$ is the inverse of the function 
$D(\delta,1)$ which is the composition of the increasing functions $(1/2)[\gamma - \log (1\!+\!\gamma/2)]$ and $\gamma=\sqrt{1+4\delta^2} -1$ previously discussed, beginning in Section 2.  This $G(r)$ is near $\sqrt {2r}$ for small $r$.  When $G(r) \!<\! (1\!-\!\rho_\alpha^2)^{1/2}/\rho_\alpha^2$ the condition 
$\lambda\!<\!1$ is satisfied and $\Delta_\alpha^{\min}=\Delta_\alpha^{target}$ indeed solves the above equation; otherwise $\Delta_\alpha^{\min} = r_\alpha \!-\! (1/2) \log \rho_\alpha^2$ provides the solution.

Now $r_\alpha = (R/a)(\log {L \choose \alpha L})/(L\log L)$. With $\alpha L$ restricted to integers between $0$ and $L$, it is not more than
$(R/a)\alpha$ and $(R/a)(1\!-\!\alpha)$, with equality at particular $\alpha$ near $0$ and $1$, respectively. It remains small, with $r_\alpha \le (R/a) (\log 2 )/\log L$, for $0 \le \alpha \le 1$.  Also we have $1-\rho_\alpha^2=\alpha(1\!-\!\alpha)v/(1\!+\!\alpha v)$ from Lemma 2.  Consequently, $\Delta_\alpha^{\min}$ is small for large $L$; moreover, for $\alpha$ near $0$ and $1$, 
it is of order $\alpha$ and $1-\alpha$, respectively, and via the indicated bounds, derivatives at $0$ and $1$ can be explicitly determined.

The analysis in Lemma 4 may be interpreted as determining section size rates $a$ such that the differentiable upper bounds on $\Delta_\alpha^{\min}$ are 
less than or equal to $\tilde \Delta_\alpha = C_\alpha \!-\! \alpha C$ for $0 \le \alpha \le 1$, where, noting that these quantities are $0$ at the endpoints of the interval, 
the critical section size rate is determined by matching the slopes at $\alpha = 1$.
At the other end of the interval, the
bound on the difference $\tilde \Delta_\alpha \!-\! \Delta_\alpha^{\min}$
has a strictly positive slope at $\alpha=0$, given by $\tau_v = (1/2) [v- \log(1\!+\!v)] - [2vR/a]^{1/2}$.

Recall that $\Delta_\alpha = C_\alpha - \alpha R - t_\alpha$.  For a sensible probability bound in Lemma 2, less than $1$, we need to arrange $\Delta_\alpha$ greater than $\Delta_\alpha^{\min}$.  This we can do if 
the threshold $t$ is less than $C_\alpha - \alpha R - \Delta_\alpha^{\min}$ and 
$t_\alpha$ is strictly between.

Express $\Delta_\alpha$ as the sum of $\Delta_\alpha^{\min}$, needed to cancel the combinatorial coefficient, and  $\Delta_\alpha^{extra} = C_\alpha - \alpha R - \Delta_\alpha^{\min} -t_\alpha$ which is positive.
This $\Delta_\alpha^{extra}$ arises in establishing that the main term in the probability bound is exponentially small.  It decomposes as $\Delta_\alpha^{extra} = \alpha (C\!-\!R) + (\tilde \Delta_\alpha \!-\! \Delta_\alpha^{\min}) - t_\alpha$, which reveals different regimes in the behavior of the exponent.  For high $\alpha$ what matters is the $\alpha (C\!-\!R)$ term, positive with $R \!<\! C$, and that $t_\alpha$ stays less than the gap $\alpha (C\!-\!R)$. 
For small $\alpha$, we approximate $\Delta_\alpha^{extra}$ by $\alpha [(C\!-\!R)+\tau_v] - t_\alpha.$

For moderate and small $\alpha$, having $R < C$ is not so important to the exponent, as the positivity of $\tilde \Delta_\alpha \!-\! \Delta_\alpha^{\min}$ produces a positive exponent even if $R$ matches or is slightly greater than $C$.  In this regime, 
the Lemma 1 bound is preferred, where we set $\Delta_\alpha = C_\alpha - \alpha R - t$ without need for $t_\alpha$.  


\section{Confirming Exponentially Small Probability}
\label{sec:conprob}
In this section we put the above conclusions together to demonstrate the reliability of approximate least squares.  The probability of the event of more than any small positive fraction of mistakes  $\alpha_0=\ell_0/L$ is shown to be exponentially small.

Recall the setting that
we have a random dictionary $X$ of $L$ sections, each of size $B$.  The mapping from $K$-bit input strings $u$ to coefficient vectors $\beta(u)$ is as previously described.  The set $\mathcal{B}$ of such vectors $\beta$ are those that have one non-zero coefficient in each section (with possible freedom for the choice of sign) and magnitude of the non-zero coefficient equal to $1$.  Let $\beta^*=\beta(u^*)$ be the coefficient vector for an arbitrary input $u^*$. We treat both the case of a fixed input, and the case that the input is drawn at random from the set of 
possible inputs. 
The codeword sent $X\beta^*$ 
is the superposition of a subset 
of terms with one from each section. 
The received string is $Y=X\beta^*+\varepsilon$
with $\varepsilon$ distributed normal $N(0,\sigma^2 I)$. The columns of $X$ are independent $N(0,(P/L)I)$ and $X$ and $Y$ are known to the receiver, but not $\beta^*$.  The section size rate $a$ is such that $B=L^a$. 

In fashion with Shannon theory, the expectations in the following theorem are taken with respect to the distribution of the design $X$ as well as with respect to the distribution of the noise; implications for random individual dictionaries $X$ are discussed after the proof.

The estimator 
$\hat \beta$ is assumed to be an (approximate) least squares estimator, taking values in $\mathcal{B}$
and satisfying $|Y\!-\!X \hat \beta|^2$ $\le |Y\!-\!X\beta^*|^2 + \delta_0$,
with $\delta_0 \ge 0$.
Let $mistakes$ denote the number of mistakes, that is, the number of sections in which the non-zero term in $\hat \beta$ 
is different from the term in $\beta^*$.

Suppose the threshold $t =\frac{\delta_0}{2\sigma} \log e$
is not more than $(1/2) \min_{\alpha \ge \alpha_0} \{\alpha (C\!-\!R) + ({\tilde \Delta}_\alpha \!-\! \Delta_\alpha^{\min})\}$.
Some natural choices for the threshold include $t=0$,
$\: t= (1/2)\alpha_0 (C\!-\!R)$, and $t= (1/2) \alpha_0 \, \tau_v$.
For positive $x$ let $g(x)=\min\{x,x^2\}$.

\vspace{.3cm}
\noindent
{\bf Theorem 5:}
Suppose the section size rate $a$ is at least $a_{v,L}$, that 
the communication rate $R$ is less than the capacity $C$ with codeword length $n= (1/R) a L \log L$, and that we have an approximate least squares estimator. 
For $\ell_0$ between $1$ and $L$, the probability $\PP[mistakes \ge \ell_0]$ is
bounded by the sum over integers $\ell$ from $\ell_0$ to $L$ of $\PP[E_\ell]$ using the minimum of the bounds from Lemmas 1 and 2. It follows that there is a positive constant $c$, such that for all $\alpha_0$ between $0$ and $1$,
$$\PP[mistakes \ge \alpha_0 L] \le 2L \exp\{-nc\min\{\alpha_0,g(C\!-\!R)\}\}.$$
Consequently, asymptotically, taking $\alpha_0$ of the order of a constant times $1/L$, the fraction of mistakes is of order $1/L$ in probability, provided $C\!-\!R$ is at least a constant multiple of $1/\sqrt L$. Moreover, for any fixed $\alpha_0$, $a$, and $R$, not depending on $L$, satisfying
$\alpha_0 \!>\! 0$, $a \!>\! a_v$ and $R \!<\! C$, we conclude that this probability is exponentially small.

\vspace{0.3cm}
\noindent{\bf Proof:}
Consider the exponent $D_{\alpha,v}=D_1(\Delta_\alpha,1\!-\!\rho_\alpha^2)$ as given at the start of the preceding section.  We take a reference $\Delta_\alpha^{ref}$ for which $\Delta_\alpha > \Delta_\alpha^{ref}$ and for which $\Delta_\alpha^{ref}$ is at least $\Delta_\alpha^{\min}$ and at least a multiple of $\tilde \Delta_\alpha$.

The simplest choice is 
$\Delta_\alpha^{ref}= \tilde \Delta_\alpha$, which may be used when $t$ is less than a fixed fraction of $\alpha_0 (C\!-\!R)$.  Then
$\Delta_\alpha = \tilde \Delta_\alpha + \alpha (C\!-\!R) - t_\alpha$ exceeds $\tilde \Delta_\alpha$, taking $t_\alpha$ to be between $t$ and $\alpha (C\!-\!R)$.  
Small precision $t$ makes for a greater computational challenge.
Allowance is made for
a more relaxed requirement that $t$
be less than $\min_{\alpha_0 \le \alpha \le 1} \{\alpha (C\!-\!R) +(1/2)\tilde \Delta_\alpha\}$ and less than a fixed fraction of $\min_{\alpha_0 \le \alpha \le 1} \{\alpha (C\!-\!R) + \tilde \Delta_\alpha \!-\! \Delta_\alpha^{\min} \}$.  Both of these conditions are satisfied when $t$ is less than the value $(1/2) \min_{\alpha \ge \alpha_0} \{\alpha (C\!-\!R) + ({\tilde \Delta}_\alpha \!-\! \Delta_\alpha^{\min})\}$ stated for the theorem.

Accordingly, set $\Delta_\alpha^{ref} = (1/2) [\Delta_\alpha + \Delta_\alpha^{min}]$ to be half way between $\Delta_\alpha^{\min}$ and $\Delta_\alpha$.  With $t$ less than both $[\alpha (C-R) + \tilde \Delta_\alpha - \Delta_\alpha^{\min}]$
and $[\alpha (C-R) + (1/2) \tilde \Delta_\alpha]$, arrange $t_\alpha > t$ to be less than both of these as well. For then $\Delta_\alpha^{ref}$ exceeds both $\Delta_\alpha^{\min}$ and $(1/4) \tilde \Delta_\alpha$ as required.

Now $D_1 (\Delta,1\!-\!\rho^2)$ has a nondecreasing derivative with respect to $\Delta$.
So $D_{\alpha,v}=D_1(\dalp,1\!-\!\rho_\alpha^2)$ is greater than $ D_{\alpha,v}^{ref}=D_1(\Delta_\alpha^{ref},1\!-\!\rho_\alpha^2)$.
Consequently, it lies above the tangent line (the first order Taylor expansion) at $\Delta_\alpha^{ref}$, that is,
$$D_{\alpha,v} \, \ge \, D_{\alpha,v}^{ref} \; + \;\; (\Delta_\alpha - \Delta_\alpha^{ref}) \,\, D^\prime,$$
where $D^\prime\!=\!D_1^\prime(\Delta)$
is the derivative of $D_1(\Delta)\!=\!D_1(\Delta,1\!-\!\rho_\alpha^2)$ with respect to $\Delta$, which is here evaluated at ${\Delta}_\alpha^{ref}$.  In detail, the derivative $D_1^\prime (\Delta)$
is seen to equal
$$\frac{1}{1+\sqrt{1\!+\!4\Delta^2/(1\!-\rho_\alpha^2)}} \, \frac{2\Delta}{1\!-\!\rho_\alpha^2}$$
when $\Delta \!<\! (1\!-\!\rho_\alpha^2)/\rho_\alpha^2$, and this derivative is equal to $1$ otherwise.
[The latter case with derivative equal to $1$ includes the situations $\alpha\!=\!0$ and $\alpha\!=\!1$ where $1\!-\!\rho_\alpha^2\!=\!0$ with $D_1\!=\!\Delta$; all other $\alpha$ have $1\!-\!\rho_\alpha^2 \!>\! 0$.]


Now lower bound the components of this tangent line. First lower bound 
the derivative $D^\prime=D_1^\prime (\Delta)$ evaluated at $\Delta=\Delta_\alpha^{ref}$.  Since this derivative is non-decreasing it is at least as large as the value at $\Delta=(1/4)\tilde \Delta_\alpha$.
As in our developments in previous sections 
$\tilde \Delta_\alpha^2 /(1\!-\!\rho_\alpha^2)$ is a bounded function of $\alpha$. Moreover, $\tilde \Delta_\alpha$ and $1\!-\!\rho_{\alpha}^2$ are positive functions of order $\alpha(1\!-\!\alpha)$ in the unit interval, with ratio tending to positive values as $\alpha$ tends to $0$ and $1$, 
so 
their ratio is uniformly bounded away from $0$.  Consequently $w_v=\min_{\alpha} D_1^\prime (\Delta_\alpha^{ref})$ is strictly positive.  [This is where we have taken advantage of $\Delta_\alpha^{ref}$ being at least a multiple of $\tilde \Delta_\alpha$; if instead we used $\Delta_\alpha^{\min}$ as the reference, then for some $\alpha$ we would find the $D_1^\prime(\Delta_\alpha^{min})$ being of order $1/\sqrt {\log L}$, producing a slightly inferior order in the exponent of the probability bound.]

Next examine $D_{\alpha,v}^{ref}$. Since $\Delta_\alpha^{ref}$ is at least $\Delta_\alpha^{\min}$, it follows 
that $D_{\alpha,v}^{ref}$ is at least $D_{\alpha,v}^{\min}=D(\Delta_\alpha^{\min},1\!-\!\rho_\alpha^2)$.

Now we are in position to apply Lemma 2 and Lemma 4. If the section size rate $a$ is at least $a_{v,L}$ we have that $nD_{\alpha,v}^{\min}$ cancels the combinatorial coefficient and hence the first term in the $\PP[E_\ell]$ bound (the part controlling $\PP[\tilde E_\ell]$) is not more than
$$\exp\{-n [\Delta_\alpha - \Delta_\alpha^{ref}] \, D^\prime\},$$
where $\alpha = \ell/L$. In the first case, with $t < \alpha (C\!-\!R)$ and $\Delta_\alpha^{ref}= \tilde \Delta_\alpha$, this yields $\PP[E_\ell]$ not more than the sum of
$$\exp\{-n[\alpha(C\!-\!R)- t_\alpha] \, D^\prime \}$$
and
$$\exp\{-n D(t_\alpha - t,\alpha^2 v/(1+\alpha^2 v))\},$$
for any choice of $t_\alpha$ between $t$ and $\alpha (C\!-\!R)$. For instance one may choose $t_\alpha$ to be half way between $t$ and $\alpha(C\!-\!R)$.

Now if $t$ is less than a fixed fraction of $\alpha_0 (C\!-\!R)$, we have arranged for both $\alpha (C\!-\!R) - t_\alpha$ and $t_\alpha - t$ to be of order $\alpha (C\!-\!R)$ uniformly for $\alpha \ge \alpha_0$.

Accordingly, the first of the two parts in the bound has exponent exceeding a quantity of order $\alpha_0 (C\!-\!R)$.  The second of the two parts has exponent related to a function of the ratio $u=(\alpha (C\!-\!R))^2/[\alpha^2 v/(1+\alpha^2v)]$
as explained in Section II, where the function is of order $u$ for small $u$ and order $\sqrt u$ for large $u$.  Here $u$ is of order $(C\!-\!R)^2$ uniformly in $\alpha$.  It follows that there is a constant $c$ (depending on $v$) such that
$$\PP[E_{\ell}] \le 2 \exp \{-nc \min\{\alpha_0 (C\!-\!R),g(C\!-\!R)\}\}.$$

An improved bound is obtained, along with allowance of a larger threshold $t$,
using $\Delta_\alpha^{ref}$ half way between $\Delta_\alpha^{min}$ and $\Delta_\alpha$. Then the first part of the bound becomes
$$\exp\{-n(1/2)[\alpha(C\!-\!R)- (\tilde \Delta_\alpha \!-\! \Delta_\alpha^{min}) - t_\alpha] \, D^\prime \}$$
provided $t_\alpha$ is chosen between $t$ and $\alpha (C\!-\!R)+ (\tilde \Delta_\alpha \!-\! \Delta_\alpha^{\min})$, e.g. half way between works for our purposes. This bound is superior to the previous one, 
when $R$ closely matches $C$, because of the addition of the non-negative $(\tilde \Delta_\alpha - \Delta_\alpha^{min})$ term.  For $\alpha$ less than, say, $1/2$, we use that the exponent exceeds a fixed multiple of $\alpha_0 \tau_v w_v$; whereas for $\alpha \ge 1/2$ we use that the exponent exceeds a fixed multiple of $(C\!-\!R) w_v$. 
For $R\!<\!C$,
it yields the desired bounds on $\PP[E_\ell]$, uniformly exponentially small for $\alpha \ge \alpha_0$, with the stated conditions on 
$t$.


With optimized $t_\alpha$, let $D_{\min,\alpha,v}$ be the minimum of the two exponents from the two terms in the bound on $\PP[E_\ell]$ at $\alpha = \ell/L$.  Likewise, let $D_{\min} =D_{\min,v}$ be the minimum of these exponents for $\ell \ge \alpha_0 L$.  We have established that $D_{\min}$ exceeds a quantity of order $\min\{\alpha_0,g(C\!-\!R)\}$.
Then for $\ell \ge \alpha_0 L$, $$\PP[E_\ell] \, \le \, 2e^{-nD_{\min}}$$
and accordingly $$\PP[mistakes \ge \alpha_0 L] \, \le \, 2L e^{-n D_{\min}}.$$

Using the form of the constants identified above, we see that even for $\alpha_0$ of order $1/L$, that is, for $\ell_0 \!=\! \alpha_0 L$ constant, the probability $\PP[mistakes \!\ge\! \ell_0]$ goes to zero polynomially in $1/L$.  Indeed, for $C\!-\!R$ at least a multiple of $1/\sqrt L$, and sufficiently small $t$, the bound becomes $2L \exp\{-n(1/2)\tau_v w_v \ell_0/L\}$ which with $n\!=\!(a/R) L \log L$ becomes,
$$\PP[mistakes \ge \ell_0] \le 2(1/L)^{(1/2)(a/R)\tau_v w_v \ell_0 - 1}.$$
It is assured to go to zero with $L$ for $\ell_0$ at least $2C/[a_v\tau_vw_v]$.
This completes the proof of Theorem 5.

\vspace{0.3cm}
\noindent
{\bf Remarks:}
For a range of values of $\ell_0$, up to the point where a multiple of $\ell_0/L$ hits $g(C\!-\!R)$, the upper tail of the distribution of the number of mistakes past a minimal value is shown to be less than that of a geometric random variable.  Using the geometric sum, an alternative to the factor $L$ outside the exponent can be arranged.



The form given for the exponential bound is meant only to reveal the general character of what is available.  In particular, via appeal to the section size analysis, we ensure to have canceled the combinatorial coefficient and yet, for $R\!<\! C$, to have enough additional exponent that the probability of a fraction of at least $\alpha_0$ mistakes is exponentially small.  A compromise was made, by introduction of an inequality (the tangent bound on the exponent) to proceed most simply to this demonstration.  Now understanding that it is exponentially small, our best evaluation avoids this compromise and proceeds directly, using for each $\alpha$ the best of the bounds from Lemma 1 and Lemma 2, as it provides substantial numerical improvement.

The polynomial bound on more than a constant number of mistakes is here extracted as an aside to the exponential bound with exponent proportional to $\ell$.  One can conclude, for sufficient section size rate $a$, using $\ell_0 = 1$, that the probability of even $1$ or more mistake is polynomially small.  Polynomially small block error probability is not as impressive when by a simple device it is made considerably better.
Indeed, we have established smaller probability bounds with larger mistake thresholds $\ell_0$.  With certain such thresholds, fewer mistakes than that are guaranteed correctable by suitable outer codes; thereby yielding smaller overall block error probability.


The probability of the error event $E=\{mistakes \ge \alpha_0 L\}$ has been computed averaging over random generation of the dictionary $X$ as well as the distribution of the received sequence $Y$.  In this case the bounds apply equally to an individual input $u$ as well as with the uniform distribution on the ensemble of possible inputs.  Implications of the bounds for a randomly generated dictionary $X$ are discussed further in Appendix A.

In the next section we review basic properties of Reed Solomon codes and discusses its role in correcting any existing
section errors.

\section{From Small Fraction of Mistakes to Small Probability of Any Mistake}
\label{sec:reed}
\vspace{0.3cm}
We employ Reed-Solomon (RS) codes (\cite{ReedSoloman1960}, \cite{LinCos}) as an outer code for correcting
any remaining section mistakes.
The symbols for the RS code come from a Galois field consisting of $q$ elements denoted by $GF(q)$, with $q$ typically  taken to be of the form $2^m$. If $K_{out},\, n_{out}$ represent  message and codeword lengths respectively, then an RS code with symbols in $GF(2^m)$ and minimum distance between codewords given by $d_{RS}$ can have the following parameters:
\algg{n_{out} &= 2^m\\
      n_{out} - K_{out} &= d_{RS} - 1
}
Here 
$n_{out} - K_{out}$ gives
the number of parity check symbols added to the message to form the codeword. 
In what follows we find it convenient to  take $B$ to be equal to $2^m$ so that can view each symbol in $GF(2^m)$ as giving a number between 1 and $B$.

We now demonstrate how the RS code can be used as an outer code in conjunction with our inner superposition code,
to achieve low block error probability. For simplicity assume that $B$ is a power of 2. First consider the case when $L$ equals $B$. Taking $m = \log_{2}B$, we have that since $L$ is equal to $B$, the RS codelength becomes $L$. Thus, one can view each symbol
 as representing an index in each of the $L$ sections. The number of input symbols is then $K_{out} = L - d_{RS} + 1$, so setting $\delta = d_{RS}/L$, one sees that the outer rate $R_{out}$, equals $1 -\delta + 1/L$ which is at least $1- \delta$.

For code composition $K_{out}\log_2B$ message bits become the $K_{out}$ input symbols to the outer code. The symbols of the outer codeword, having length $L$, gives the labels of terms sent from each section using our inner superposition
with codelength $n = L\log_2B/R_{inner}$. From the received $Y$ the estimated labels $\hat j_i, \hat j_2, \ldots \hat j_L$ using our least squares decoder can be again thought of as output symbols for our RS codes. If $\hat{\delta}_e$
denotes the section mistake rate, it follows from the distance property of the outer code that if $2\hat{\delta}_e \leq \delta$ then these errors can be corrected. The overall rate $R_{comp}$ is seen to be equal to the product of rates  $R_{out}R_{inner}$ which is at least $(1 - \delta)R_{inner}$. Since we arrange for $\hat{\delta}_e$ to be smaller than some $\alpha_0$ with exponentially small probability, it follows from the above that composition with an outer code
allows us to communicate with the same reliability, albeit with a slightly smaller rate given by $(1- 2\alpha_0)R_{inner}$.

The case when $L < B$ can be dealt with by observing (\cite{LinCos}, Page 240) that an $(n_{out},K_{out})$ RS code  as above, can be shortened by length $w$, where $0\leq w < K_{out}$, to form an $(n_{out} - w, K_{out} - w)$ code with the same minimum distance $d_{RS}$ as before. This is easily seen by viewing each codeword as being created by appending $n_{out} - K_{out}$ parity check symbols to the end of the corresponding message string. Then the code formed by considering the set of codewords with the $w$ leading symbols identical to zero has precisely the properties stated above. 

With $B$ equal to $2^m$ as before, we have $n_{out}$ equals $B$ so taking $w$ to be $B - L$ we
get an $(n_{out}', K_{out}')$ code, with $n_{out}' = L$, $K_{out}' = L - d_{RS} + 1$ and minimum distance $d_{RS}$.
Now since the codelength is $L$  and symbols of this code are in $GF(B)$ the code composition can be carried out as before.

We summarize the above in the following.

\vspace{0.3cm}
\noindent {\bf Proposition 6:} To obtain a code with small block error probability it is enough to have demonstrated a partitioned superposition code for which the section error rate is small with high probability. In particular, for any given positive $\epsilon$ and $\alpha_0$, let $R$ be a rate for which the partitioned superposition code with $L$ sections has
$$\mbox{Prob}\{ \#\, section\,\, mistakes > \alpha_0 L \} \leq  \epsilon.$$
Then through concatenation of such a code with an outer Reed-Solomon code, one obtains a composite code for which the rate is $(1 - 2\alpha_0)R$ and the block error probability is less than or equal to $\epsilon$.

\section*{Appendix A: Implications for random dictionaries}
Here we provide discussion of the implications of our error probability bound of Section V for randomly generated dictionaries $X$.

The probability of the error event $E=\{mistakes\ge \alpha_0 L\}$ has been computed averaging over random generation of the dictionary $X$ as well as the distribution of the received sequence $Y$.  Let's denote the given bound $P_{e}$.
The theorem asserts that
this bound is exponentially small.  For instance, it is less than $2L e^{-nD_{min}}$.

The same bound holds 
for any given $K$ bit input sequence $u$.
Indeed, the probability of $E$ given that $u$ is sent, which we may write as
$\PP[E|u]$ is the same for all $u$ by exchangeability of the distribution of the columns of $X$. 
Accordingly, it also matches the average probability ${ {\PP}}[E] = \frac{1}{2^K} \sum_u \PP[E|u]$, averaging over all possible inputs, so this average probability will have the same bound.

Reversing the order of the average over $u$ and the average over the choice of dictionary $X$, the average probability may be written $\E_X \big[\frac{1}{2^K} \sum_u \PP[E|u, X]\big]$, 
where $\PP[E|u,X]$ denotes the probability of the error event $E$, conditioning on the event that the input is $u$ and that the dictionary is $X$ (the only remaining average in $\PP[E|u,X]$ is over the distribution of the noise).  This $\PP[E|u,X]$ will vary with $u$ as well as with $X$.

An appropriate target performance measure is
$$\PP[E|X]=\frac{1}{2^K} \sum_u \PP[E|u, X],$$
the probability of the error event, averaged with respect to the input, conditional on the random dictionary $X$.  Since the expectation $\PP[E]=\E \big[ \PP[E|X] \big]$ satisfies the indicated bound, random $X$ are likely to behave similarly.  Indeed, by Markov's inequality $\PP\big[ \PP[E|X] \ge \tau P_e^{b}\big] < 1/\tau$.

So with a single draw of the dictionary $X$, it will satisfy $\PP[E|X] \le \tau P_{e}^{b}$, with probability at least $1-1/\tau$.  The manageable size of the dictionary facilitates computational verification by simulation that the bound holds for that $X$.  With $\tau=2$ one may independently repeat the generation of $X$ a Geometric($1/2$) number of times until success.  The mean number of draws of the dictionary required for one with the desired performance level is $2$.  Even with only one draw of $X$, one has with $\tau= e^{(n/2)D_{min}}$, that
$$\PP[E|X] \le 2L e^{-(n/2)D_{min}},$$ except for $X$ in an event of probability not more than $e^{-(n/2)D_{min}}$.

Now $\PP[E|X]$ exponentially small implies that $\PP[E|u,X]$ is exponentially small for most $u$ (again by Markov's inequality). In theory one could expurgate the codebook, leaving only good performing $\beta$ and reassigning the mapping from $u$ to $\beta$, to remove the minority of cases in which $\PP[E|u,X] > 4L e^{-(n/2)D_{min}}$. Thereby one would have uniformly exponentially small error probability.

In principle, simulations can be used to evaluate $\PP[E|u,X]$ for a specific $\beta$ and $X$, to decide whether that $\beta$ should be used.  However, it is not practical to do so in advance for all $\beta$, and it is not apparent how to perform such expurgations efficiently on-line during communications.
Thus we maintain our focus in this paper on average case error probability, averaging over the possible inputs, rather than maximal error probability.

As we have said, for the average case analysis, armed with a suitable decoder, one can check, for a dictionary $X$, whether it satisfies an exponential bound on $\PP[E|X]$ empirically by simulating a number of draws of the input and of the noise.
Nevertheless, it would be nice to have a more direct, non-sampling check that a dictionary $X$ satisfies requirement for such a bound on $\PP[E|X]$.
Our current method of proof does not facilitate providing such a direct check.  The reason is that our analysis does not exclusively use the distribution of $Y$ given $u$ and $X$; rather it makes critical use of properties of the joint distribution of $Y$ and $X$ given $u$.

Likewise, averaging over the random generation of the dictionary, permits a simple look at the satisfaction of the average power constraints.
With a randomly drawn $u$, and associated coefficient vector $\beta=\beta(u)$, consider the behavior of the power $|X\beta|^2$ and whether it stays less than $(1\!+\!\epsilon)P$. The event $A^c =\{|X\beta|^2 \!\ge\! (1\!+\!\epsilon)P\}$, when conditioning on the input $u$,
has exponentially small probability $\PP[A^c|u]$, in accordance with the normal distribution of the codeword obtained via the distribution of the dictionary $X$.  Again $\PP[A^c|u]$ is the same for all $u$ and hence matches the average
${\bar {\PP}}[A^c]$ with expectation taken with respect to random input $u$ as well as with respect to the distribution of $X$. So reversing the order of the expectation we have that $\E\big[\PP[A^c|X]\big]$ enjoys the exponential bound, from which, again by applications of Markov's inequality, except for $X$ in an event of exponentially small probability, $|X\beta|^2 < (1+\epsilon)P$ for all but an exponentially small fraction of coefficient vectors $\beta$ in $\mathcal{B}$.

Control of the average power is a case in which we can formulate a direct check of what is required of the dictionary $X$, as is examined in Appendix B.

\section*{Appendix B: Codeword power}

Here we examine the average and maximal power of the codewords.  The maximal power has a role in our analysis of decoding.

The power of a codeword $c$ is its squared norm $|c|^2$, consisting of the average square of the codeword values across its $n$ coordinates.
The terminology {\it power} arises from settings in which codeword values are voltages on a communication wire or a transmission antenna in the wireless case, recalling that power equals average squared voltage divided by resistance.

\vspace{0.3cm}
\noindent
{\bf Average power for the signed subset code:} Consider first our signed, subset superposition code. Each input correspond to a coefficient vector $\beta=(\beta_j)_{j=1}^N$, where for each of the $L$ sections there is only one $j$ for which $\beta_j$ is nonzero, and, having absorbed the size of the terms into the $X_j$, the nonzero coefficients are taken to be $\pm 1$.  These are the 
coefficient vectors $\beta$ of our codewords 
$c=X\beta$, for which the power is $|c|^2=|X\beta|^2$.

With a uniform distribution on the binary input sequence of length $K= L\log (2B)$, the induced distribution on the sequence of indices $j_i$ is independent uniform on the $B$ choices in section $i$, and likewise the signs 
are independent uniform $\pm 1$ valued, for $i=1,2,\ldots,L$.  Fix a dictionary $X$, and consider the average of the codeword powers with this uniform distribution on inputs,
$${\bar P}_X = \frac{1}{2^K} \sum_{\beta} |X\beta|^2.$$
By independence across sections, this average 
simplifies to
$${\bar P}_X = \sum_{i=1}^L \frac{\sum_{j \in sec_i} |X_j|^2}{B}.$$
Now we consider the size of this average power, using the distribution of the dictionary $X$, with each entry independent Normal($0,P/L$).  This average power ${\bar P}_X$ has mean $\E{\bar P}_X$ equal to $P$,
standard deviation $P\sqrt {2/(Nn)}$, and distribution equal to $[P/(Nn)] {\Xcal}_{Nn}^2$, where ${\Xcal}_{d}^2$ is a Chi-square random variable with
$d=Nn$ degrees of freedom.  Accordingly $\bar P_X$ is very close to $P$.

Indeed, in a random draw of the dictionary $X$, the chance that $\bar P_X$ exceeds $P + 2 P  \sqrt{{(\log(1/\epsilon))}{/(Nn)}}$ is approximately less than $\epsilon$, as can be seen via the Chernoff-Cramer bound $P\{\Xcal_d^2 > d + a \sqrt {2d}\} \le e^{-dD_2(a\sqrt{2/d})}$, for positive $a$, where the exponent $D_2(\delta) = (1/2)[\delta - \log(1\!+\!\delta)]$ is near $\delta^2/4$ for small positive $\delta$, so that the bound is near $e^{-a^2/2}$, which is $\epsilon$ for $a=\sqrt{2\log(1/\epsilon)}$.

Or we may appeal to the normal approximation for fixed $a$ when $d=Nn$ is large; the probability is not more than $0.05$ that the dictionary has average power $\bar P_X$ outside the interval formed by the mean plus or minus two standard deviations
$$P \, \pm \, 2P\sqrt{2/(Nn)}.$$

For instance, suppose $P=15$ and the rate is near the capacity $C=2$, so that $Nn$ is near $(LB)(L\log B)/C$, and pick $L=64$ and $B=256$. Then with high probability $\bar P_X$ is not more than $1.001$ times $P$.

If the average power constraint is held stringently, with average power to be precisely not more than $P$, then in the design of the code proceed by generating the entries of $X$ with power $P'/L$, where $P'$ is less than $P$.  The analysis of the preceding sections then carries through to show exponentially small probability of more than a small fraction of mistakes when $R<C$ as long as $P'$ is sufficiently close to $P$.

\vspace{0.3cm}
\noindent
{\bf Average power for the subset code:} Likewise, let's consider the case of subset superposition coding without use of the signs.  Once again fix $X$ and consider a uniform distribution on inputs; it again makes the term selections $j_i$ independent and uniformly distributed over the $B$ choices in each section. Now there is a small, but non-zero, average $\bar X_i = (1/B)\sum_{j \in sec_i} X_j$ of the terms in each section $i$, and likewise a very small, but non-zero, overall average $\bar X = (1/L) \sum_{i=1}^L \bar X_i$.  We need to make adjustments by these averages when invoking the section independence to compute the average power.  Indeed, as in the rule that an expected square is the square of the expectation plus a variance, the average power is the squared norm of the average of the codewords plus the average norm squared difference between codewords and their mean. The mean of the codewords, with the uniform distribution on inputs, is $\sum_{i=1}^L \bar X_i = L \bar X$, which is a Normal($0,(P/B)I$) random vector of length $n$.

By independence of the term selections, the codeword variance is $\sum_{i=1}^L (1/B) \sum_{j \in sec_i} |X_j - \bar X_i|^2$.  Accordingly, in this subset coding setting, $${\bar P}_X =\sum_{i=1}^L \frac{\sum_{j \in sec_i} |X_j- \bar X_i|^2}{B} + |\sum_{i=1}^L \bar X_i|^2.$$
Using the independence of $\bar X_i$ and $(X_j- \bar X_i: j \in sec_i)$ and standard distribution theory for sample variances,  with a randomly drawn dictionary $X$, we have that ${\bar P}_X$ is  $P/(LBn)$ times a Chi-square random variable with $nL(B\!-\!1)$ degrees of freedom, plus $P/(nB)$ times an independent Chi-square random variable with $n$ degrees of freedom.  So it has mean equal to $P$ and a standard deviation of $P \sqrt{ \frac {2}{n}} \sqrt {\frac{1}{LB} + \frac{1-1/L}{B^2}}$, which is slightly greater than before.  It again yields only a small departure from the target average power $P$, as long as $n$ and $B$ are large.

\vspace{0.3cm}
\noindent
{\bf Worst case power:}
Next we consider the matter of the size of the maximum power $P_X^{max} = \max_{\beta} |X\beta|^2$ among codewords for a given design $X$.  The simplest distribution bound is to note that for each $\beta$, the codeword $X\beta$ is distributed as a random vector with independent Normal($0,P$) coordinates, for which $|X\beta|^2$ is $P/n$ times a Chi-square $n$ random vector.  There are $e^{nR}$ such codewords, with the rate written in nats. We recall the probability bound $\PP\{\Xcal_n^2 \!>\! n(1\!+\!\delta)\} \le e^{-n D_2(\delta)}$. Accordingly, by the union bound, $P_X^{max}$ is not more than $$P \, +\,  P \,\, G_2 \left(\,R + \frac{1}{n}{\log (1/\epsilon)}\right)$$ except in an event of probability which we bound by $e^{nR} e^{-nD_2(G_2(R+ (\log 1/\epsilon)/n))}= \epsilon$, where $G_2$ is the inverse of the function $D_2(\delta)= (1/2) [\delta- \log(1\!+\!\delta)]$. This $G_2(r)$ is seen to be of order $2\sqrt r$ for small positive $r$ and of order $2 r$ for large $r$.  Consequently, the bound on the maximum power is near $P+P\,G_2(R)$ rather than $P$.

According to this characterization, for positive rate communication, with subset superpositions, one can not rely, either in encoding or in decoding, on the norms $|X\beta|^2$ being uniformly close to their expectation.

\vspace{0.3cm}
\noindent
{\bf Individual codeword power:} We return to signed subset coding and provide explicitly verifiable conditions on $X$ such that for every subset, the power $|X\beta|^2$ is near $P$ for most choices of signs.  The uniform distribution on choices of signs ameliorates between-section interference to produce simplified analysis of codeword power.



The input specifies the term $j_i$ in each sections along with the choice of its sign given by $\sign_i$ in $\{-1,+1\}$, leading to coefficient vectors $\beta$ equal to $\sign_i$ at position $j_i$ in section $i$, for $i=1,2,\ldots, L$.  The uniform distribution on the choices of signs leads to them being independently, equiprobable $+1$ and $-1$.

Now the codeword is given by $X\beta = \sum_{i=1}^L \sign_i \, X_{j_i}$.  It has the property that conditional on $X$ and the subset $S=\{j_i:i=1,2,\ldots,L\}$, the contributions $\sign_i \, X_{j_i}$ for distinct sections are made to be mean zero uncorrelated vectors by the random choice of signs.  In particular, again conditioning on the dictionary $X$ and the subset $S$, we have that the power $|X\beta|^2$ has conditional mean $$P_{X,S} =\sum_{i=1}^L |X_{j_i}|^2,$$ which we shall see is close to $P$.  The deviation from the conditional mean $|X\beta|^2 - P_{X,S}$ equals $\sum_{i\ne i'} \sign_i \sign_{i'} X_{j_i} \cdot X_{j_{i'}}$. The presence of the random signs approximately symmetrizes the conditional distribution and leads to conditional variance $2 \sum_{i\neq i'} (X_{j_i} \cdot X_{j_{i'}})^2$.

Now concerning the columns of the dictionary, the squared norms $|X_j|^2$ are uniformly close to $P/L$, since the number of such $N=LB$ is not exponentially large.  Indeed, by the union bound the maximum over the $N$ columns, satisfies $$\max_j |X_j|^2 \, \le \, \frac{P}{L} \,+\, \frac{P}{L} \,\, G_2\left( \frac{1}{n}\log (N/\epsilon)\right),$$ except in an event of probability bounded by $\epsilon$.

Whence the conditional mean power $P_{X,S}$ is not more than $$P \, +\, P \,\, G_2 \! \left( \frac{1}{n}\log (N/\epsilon)\right),$$ uniformly over all allowed selections of $L$ term subsets.  Note here that the polynomial size of $N=LB$ makes the $(\log N)/n$ small; this is in contrast to the worst case analysis above were the log cardinality divided by $n$ is the fixed rate $R$.

Next to show that the conditional mean captures the typical power, we show that the conditional variance is small.  Toward that end we examine the inner products $X_j \cdot X_{j'}$ and their maximum absolute value $\max_{j < j'} |X_j \cdot X_{j'}|$.  Consider products of independent standard normals $Z_1 Z_2$.  These have moment generating function $\E e^{\lambda Z_1 Z_2}$ equal to $1/(1-\lambda^2)^{1/2}$.  [This matches the moment generating function for half the difference in squares of independent normals found in Section 2; to see why note that $Z_1Z_2$ equals half the difference in squares of $(Z_1+Z_2)/\sqrt 2$ and $(Z_1-Z_2)/\sqrt 2$.]

Accordingly $\PP \{X_j \cdot X_{j'} \ge (P/L)\Delta\} \le e^{-nD(\Delta)}$, for positive $\Delta$, where $D(\Delta)=D(\Delta,1)$. As previously discussed, this $D(\Delta)$
is near $\Delta^2/2$ for small $\Delta$ and accordingly its inverse function $G(r)$ is near $\sqrt {2r}$ for small $r$. 
The corresponding two-sided bound is $\PP \{|X_j \cdot X_{j'}| \ge (P/L)\Delta\} \le 2e^{-nD(\Delta)}$. 
By the union bound, we have that
$$\max_{j < j'} |X_j \cdot X_{j'}| \le \frac{P}{L} \, G\left(\frac{1}{n}\log (N^2/\epsilon)\right),$$
except for dictionaries $X$ in an event of probability not more than $\epsilon$.

Recall that the conditional variance of $|X\beta|^2$ equals $2 \sum_{i\neq i'} (X_{j_i} \cdot X_{j_{i'}})^2$.
In the likely event that the above bound holds, we have that this conditional variance is not more than $2 P^2 G^2 (\frac{1}{n}\log(N^2/\epsilon))$. Consequently, the conditional distribution of the power $|X\beta|^2$ given $X$ and $S$ is indeed concentrated near $P$.

Accordingly, for each subset, most choices of sign produce a codeword with power $|X\beta|^2$ near $P$.  Moreover, for this codeword power property, it is enough that the individual columns of the dictionary have $|X_j|^2$ near $P/L$ and $X_j \cdot X_{j'}$ near $0$, uniformly over $j\ne j'$.


\section*{Acknowledgment}

We thank John Hartigan, Cong Huang, Yiannis Kontiyiannis, Mokshay Madiman, Xi Luo, Dan Spielman, Edmund Yeh, John Hartigan, Mokshay Madiman, Dan Spielman, Imre Teletar, Harrison Zhou, David Smalling and Creighton Heaukulani for helpful conversations.  


\end{document}